\documentclass[10pt,twocolumn]{article}

\usepackage[utf8]{inputenc}
\usepackage[T1]{fontenc}
\IfFileExists{lmodern.sty}{\usepackage{lmodern}}{}
\usepackage[english]{babel}
\usepackage[a4paper,top=2.2cm,bottom=2.4cm,left=1.8cm,right=1.8cm,columnsep=0.65cm]{geometry}
\usepackage{amsmath,amsfonts,amssymb}
\usepackage{graphicx}
\usepackage{xcolor}
\usepackage{float}
\usepackage{multirow}
\usepackage{makecell}
\usepackage{subcaption}
\usepackage[font=small,labelfont=bf]{caption}
\usepackage{authblk}
\usepackage[numbers,sort&compress]{natbib}
\usepackage{microtype}
\usepackage{titlesec}
\titleformat*{\section}{\large\bfseries}
\titleformat*{\subsection}{\normalsize\bfseries}
\titleformat*{\subsubsection}{\normalsize\bfseries\itshape}
\usepackage[colorlinks=true,allcolors=blue]{hyperref}

\title{\Large\bfseries The joint dynamics of Personal, Expressed and Social beliefs}

\author[a]{Filippo Zimmaro}
\author[b,c,d]{Mirta Galesic}
\author[b,c]{Henrik Olsson}

\affil[a]{Institute for Cognitive Sciences and Technologies, Italian National Research Council (CNR), Rome, Italy}
\affil[b]{Complexity Science Hub, Vienna, Austria}
\affil[c]{Santa Fe Institute, Santa Fe New Mexico, US}
\affil[d]{Vermont Complex Systems Institute, Burlington, Vermont, US}

\date{}

\begin{document}


\twocolumn[
\begin{@twocolumnfalse}
\maketitle
\vspace{-1.5em}
\begin{abstract}
Understanding belief dynamics is essential for dealing with societal phenomena from misperceptions and conflicts to polarization and stalemates in collective action. However, it is impeded by two main challenges. First, rich and largely qualitative knowledge about socio-cognitive processes involved in belief change and spread is insufficiently integrated into computational models of belief dynamics. Second, while numerous models of belief dynamics exist, they are rarely compared because of the differences in underlying theoretical frameworks and modeling choices. Here we address these two problems by developing a Personal, Expressed, and Social Beliefs (PES) meta-model in which personal beliefs co-evolve with one’s expressed beliefs and perceived beliefs of others. It formalizes many critical socio-cognitive processes including authenticity, self-reinforcement, ego projection, social influence, conformity, validity, context effects, inertia, and potential and felt dissonance. It also enables principled comparison of a wide range of existing belief-dynamics models as its nested versions, from Voter, Ising, and DeGroot models, to bounded confidence and belief network models. We show how PES can be used to compare possible mechanisms underlying pluralistic ignorance, and find that it can occur when people conform their expressed beliefs to perceived beliefs of others, in particular when they have limited expressive repertoire (e.g., like, dislike, or stay silent). Counterintuitively, pluralistic ignorance can be dissolved when everyone ego-projects their own personal beliefs onto others to the same extent. Ego-projection generates misperception only when it is heterogeneous and covaries with personal beliefs. These results yield testable hypotheses and help predict when interventions will change personal, expressed, or social beliefs.

\end{abstract}
\begin{center}
\begin{minipage}{0.9\textwidth}
\end{minipage}
\end{center}
\vspace{1.5em}
\end{@twocolumnfalse}
]

{\renewcommand{\thefootnote}{}\footnotetext{\raggedright Corresponding author: Filippo Zimmaro.\\ E-mail: \href{mailto:zimmarofilippo@gmail.com}{zimmarofilippo@gmail.com}}}


Belief dynamics research seeks to understand how interacting social and cognitive processes lead to changes in individual beliefs and their spread on social networks. Decades of research across the social and behavioral sciences have produced empirical findings, theories, and formal models of these processes. Research relevant to personal belief updating includes social influence \cite{cialdini2004social,brechwald2011beyond,spears2021social}, self-reinforcement through self-perception and self-deception \cite{bem1967self,cohen1962experiment,vonhippel2011evolution}, context effects arising from relationships among beliefs and external influences \cite{Festinger_1957,Heider_1958,brandt2021evaluating,dalege2023networks}, and inertia associated with dependence on previous personal beliefs \cite{ye2019influence}. Research on belief expression examines authenticity \cite{sheldon1997trait,wood2008authenticity,tourangeau2007sensitive} and conformity \cite{cialdini2004social,claidiere2012integrating}, while research on perceptions of others examines validity \cite{funder1995accuracy,robbins2005social,goel2010real,olsson2023cognitive} and ego projection \cite{ross1977false,dawes1989statistical,Epley_2004,krueger2007social}. Despite this extensive research, two challenges remain. The first challenge is integrating knowledge about these processes, which remains dispersed across disciplines, into a meta-model that allows them to be studied jointly \cite{galesic2021human}. Such an integration is needed to understand how personal beliefs, belief expression, and perceptions of others jointly change and interact to produce collective phenomena.

A second challenge is comparing numerous computational models of belief dynamics developed from different theoretical foundations. These include models grounded in social and psychological theories as well as a substantial body of work inspired by statistical physics, which develops simple computational models to predict temporal and spatial patterns in belief dynamics and reproduce phenomena such as polarization or consensus \cite{starnini2026opinion,castellano2009statistical,Flache_2017}. However, these models are still rarely compared systematically with one another and with empirical data. One important reason is that they draw on different analogies for the socio-cognitive systems underlying belief dynamics and consequently emphasize different processes \cite{olsson2024analogies}. Some model belief spread as an epidemiological process \cite{bettencourt2006power}, others draw on ferromagnetic interactions \cite{galam1997rational}, while others use thresholds for social influence \cite{deffuant2000mixing,hegselmann2002opinion}, structural balance \cite{cartwright1956structural,pham2020effect}, weighting and adding \cite{friedkin1990social,proskurnikov2017tutorial,gavrilets2021coevolution}, or Bayesian updating \cite{acemoglu2011opinion,cook2016rational}. Models can also represent network structures connecting individuals or beliefs \cite{brandt2021evaluating,dalege2023networks,goldberg2018beyond,proskurnikov2017tutorial,rodriguez2016collective}. They further differ in whether they distinguish personal beliefs from expressed beliefs, as in Concealed Voter and Expressed-Private Opinion models \cite{gastner2019impact,ye2019influence}, or personal beliefs from social beliefs, as in Networks of Beliefs theory \cite{dalege2023networks}. 

This diversity makes it difficult to map models onto one another, compare their predictions, and understand the reasons for their disagreements. It also makes it difficult to determine which socio-cognitive processes are shared across models, which are omitted, and what their inclusion would change. A particularly important issue is whether models distinguish between personal beliefs, expressed beliefs, and social beliefs about others. Equating personal and expressed beliefs rules out inauthentic expression, while equating social beliefs with others’ expressed beliefs rules out misperception arising from processes such as ego projection. These assumptions constrain which interactions among processes can be studied. For example, examining how conformity and ego projection jointly contribute to pluralistic ignorance requires allowing personal, expressed, and social beliefs to differ. Models that omit these distinctions have limited ability to represent situations in which people misperceive what others believe, express something different from what they personally believe, or update beliefs through the combined effects of these processes. This makes them less suitable for describing real-world belief dynamics and proposing interventions to reduce the likelihood of harmful trajectories.

\begin{figure*}[t!]
    \centering
    \IfFileExists{Fig1.pdf} 
    {\includegraphics[scale=0.75]{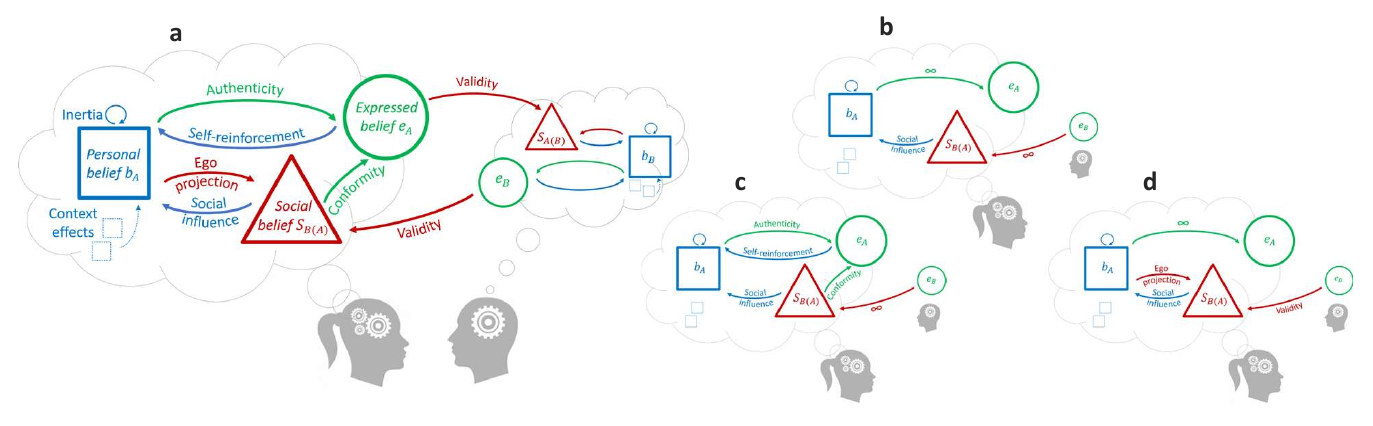}}{\fbox{\parbox[c][6cm][c]{0.82\textwidth}{\centering Placeholder for Figure 1: PES meta-model. Upload \texttt{Fig1.pdf} to display the original figure.}}}
    \caption{\textbf{PES meta-model.} Simplified representation of the structure and processes in the PES model. a) The complete PES framework. The focal individual $i$ on the left interacts with the individual $j$ on the right. Squares show personal beliefs, triangles social beliefs, and circles expressed beliefs of the individuals. Each belief is updated depending on the dissonance felt from the other beliefs through different socio-cognitive processes indicated on the edges. For example, the update of the social belief $s_{j(i)}$ is influenced by $j$'s expressed belief through the process of validity, and by $i$'s personal belief through the process of ego projection. The combination of these two processes forms the corresponding dissonance function, which guides the updates. The dashed squares represent the non-focal personal beliefs, influencing the focal personal belief through context effects. The three smaller panels show how different classes of models can be mapped on and compared through the PES framework (see text and Table \ref{tab:all_models_unified} for details). b) Class 1 models do not distinguish between belief types. c) Class 2 models distinguish between personal and expressed beliefs, but do not include social beliefs. d) Class 3 models distinguish between personal and social beliefs, but do not include expressed beliefs. }
    \label{fig:PES_model}
\end{figure*}


Here we address these two challenges by developing the Personal, Expressed, and Social Beliefs (PES) meta-model. First, PES formalizes socio-cognitive processes that have been studied extensively but are rarely represented jointly within a single belief-dynamics framework. 
Second, PES represents a range of existing belief-dynamics models within a shared formal structure, making their assumptions explicit and enabling comparison of the processes and interactions they can capture. We illustrate how these contributions work together through an application to pluralistic ignorance. By varying conformity and ego projection separately and jointly, we examine how distorted expression and biased perception produce or reduce collective misperception. This application demonstrates how the framework can be used to investigate competing explanations of the same phenomenon and develop hypotheses for empirical tests and interventions. We end by discussing how the framework contributes to the study of belief dynamics and the new research directions it opens.

\begin{table*}[t!]
\centering
\resizebox{0.8\textwidth}{!}{
\begin{tabular}{|p{2.5cm}|p{5cm}|p{7cm}|}
\hline
\textbf{Process (link)} & \textbf{Adjoining nodes} & \textbf{Description} \\ \hline \hline
Social Influence & Personal belief $\leftarrow$ Social beliefs & Social beliefs influence personal belief through persuasion, compliance, socialization, peer pressure, imitation, etc.\\ \hline
Self-Reinforcement & Personal belief $\leftarrow$ Expressed belief & Personal belief is reinforced by expressed belief through self-perception or self-deception. \\ \hline
Context effects & Personal belief $\leftarrow$ Non-focal Personal beliefs / External influence & 
Related beliefs in one’s belief network, or external inputs such as new information, media reports, or transformative events, influence the focal personal belief. \\ \hline
Inertia & Personal belief $\leftarrow$ Past personal belief& The past personal belief influences the update of the personal belief. \\ \hline
\hline
Authenticity & Expressed belief $\leftarrow$ Personal belief  & 
Personal belief influences expressed belief because of personal preferences or strategic considerations \\ \hline
Conformity &  Expressed belief $\leftarrow$ Social beliefs & 
Social beliefs influence expressed belief because of informational or normative motivations. \\ \hline
\hline
Validity & Social belief $\leftarrow$ Expressed belief of others & Beliefs expressed by others tend to be reflected in one's perceptions of others' beliefs, with or without distortions. \\ \hline
Ego Projection & Social belief $\leftarrow$ Personal belief  & Personal belief about an issue is projected into the representation of others' beliefs about that issue. \\ \hline
\end{tabular}
}
\caption{\textbf{Socio-cognitive processes within the PES model.} Processes influencing personal beliefs (top), expressed beliefs (middle), and social beliefs (bottom).}
\label{table:processes_links}
\end{table*}

\section*{Socio-Cognitive Processes in the PES Framework  } 


The distinction between personal beliefs, expressed beliefs, and social beliefs allows the PES model to represent a range of socio-cognitive processes through directed connections between belief types (Figure \ref{fig:PES_model}, Table \ref{table:processes_links}). Building on social psychological research on dissonance and ambivalence \cite{Dhami_2008, Festinger_1957, Gawronski_2012, Gawronski_2012b, Heider_1946, Heider_1958, Newbyclark_2002}, the PES model differentiates between \textit{potential dissonance}, defined as inconsistencies within or between beliefs, and \textit{felt dissonance}, defined as the psychological discomfort that arises when individuals attend to these inconsistencies. When individuals pay attention to these potential dissonances, they become felt and can influence belief updating \cite{dalege2023networks}. 

\subsection*{Updating Personal Beliefs}
Personal beliefs are updated through reducing dissonance with social beliefs, expressed beliefs, other personal beliefs, and past personal beliefs. Each source of influence corresponds to a distinct socio-cognitive process.

\textbf{Social Influence:} Social influence can be a powerful determinant of belief change and it can originate from many sources ranging from one's peers \cite{brechwald2011beyond} to the norms of the group one belongs to \cite{miller2016changing, spears2021social}. In the PES model, social influence is represented as a directed link from social beliefs to the focal individual's personal belief. Social influence processes include persuasion, compliance, obedience, socialization, peer pressure, and imitation \cite{cialdini2004social}. In contrast to most network approaches and models of social influence \cite{Flache_2017}, we do not assume a direct influence of others' beliefs on personal beliefs; instead, influence occurs through individuals' perceptions of others' beliefs (social beliefs). This is an important distinction, as in order for the expressed beliefs of others to influence personal beliefs, they must be perceived and represented by an individual through their social beliefs \cite{thomas1928,galesic2021human}.

\textbf{Self-Reinforcement:} Individuals' personal beliefs can be influenced by their own expressed beliefs. In the PES model, self-reinforcement is represented as a directed link from the expressed belief to the personal belief. Individuals often develop or adjust their beliefs by observing their own behaviors. Once individuals express a belief, they will be more likely to infer that this is their actual personal belief \cite{bem1967self}. For example, expressing political beliefs on social media reinforces partisan thinking \cite{cho2018influencing}, and experiments with counterattitudinal essay writing suggest that personal beliefs can be influenced by expressing beliefs that are against one's own \cite{cohen1962experiment, linder1967decision}. Self-reinforcement is also closely linked to self-deception, which may have evolved to support interpersonal deception by helping individuals avoid conscious cues that could reveal their intent to deceive \cite{vonhippel2011evolution}. 

\textbf{Context Effects:} Personal beliefs are not only directly influenced by their connections to an individual's expressed and social beliefs, but also by belief context. In the PES model, the influence of belief context can be divided into two separate processes. First, inconsistency between personal beliefs can cause personal dissonance, and this can influence the change of personal beliefs \cite{dalege2023networks}. For example, believing that vaccines are effective at preventing diseases but also dangerous is inconsistent. Second, belief context can also represent influences from external sources, such as new information, media sources, transformative events, or the fitness benefits or drawbacks of different beliefs. In statistical physics models, this is represented as an external field that biases a belief node \cite{brandt2021evaluating, dalege2018attitudinal, yildiz2013binary}.

\textbf{Inertia:} People might not immediately change their beliefs due to path dependencies, memory effects, risk aversion or other factors that reduce the possibility of jumps to new belief states. In the PES model, we implement inertia for personal beliefs as a memory process that depends on previous belief states. The same mechanism could also represent persistence in social beliefs or reluctance to change expressed beliefs in future extensions. Assumptions of this kind of inertia are routinely incorporated in cognitive models of subjective probability \cite{nilsson2005cognitive}, decision-making models \cite{gonzalez2011instance}, reinforcement learning \cite{erev2010choice}, cultural evolution \cite{henrich2002modeling}, and Bayesian priors \cite{lewandowsky2019influence}.

\subsection*{Updating Expressed Beliefs}
Expressed beliefs are updated through reducing dissonance with personal beliefs and social beliefs, corresponding to the processes of authenticity and conformity.

\textbf{Authenticity:} People generally want to express their true self, and consequently, they have a preference for authenticity \cite{brown2022social}. In the PES model, this is represented by a directed link from personal beliefs to expressed beliefs. Authenticity can be conceptualized as consistency across roles or as psychological authenticity within each role \cite{sheldon1997trait}. There are, however, reasons for people to not always be authentic. Expressed beliefs can differ from personal beliefs to gain social approval, avoid being ostracized \cite{cialdini2004social}, or to influence and manipulate others for personal gain \cite{henrich2009evolution}. The issue of authenticity is also a key methodological concern, where socially desirable responding can undermine the validity of self-reports \cite{tourangeau2007sensitive}.

\textbf{Conformity:} Conformity is defined as the act of adjusting one's perceptions, beliefs, or behaviors to align with those of others. In the PES model, conformity is represented by directed links between all the social beliefs related to the interacting individuals and the expressed belief. Conformity can be driven either by informational motivations to form an accurate interpretation of reality or by normative motivations to obtain social approvals of others \cite{claidiere2012integrating}. Anti-conformity instead involves responding in opposition to others' positions \cite{willis1965conformity,nyczka2013anticonformity}. Applied to expressed beliefs, it concerns expressing disagreement with perceived others. 

\subsection*{Updating Social Beliefs}
Social beliefs are formed based on the dissonance with expressed beliefs of others and one's own personal beliefs.

\textbf{Validity:} The social belief of individuals is affected by the interacting party's expressed belief. In the PES model, this represents the validity of the perception of another person's expressed belief. Extensive research investigates how aligned or accurate are people's perceptions of others' beliefs \cite{bernard1984problem, freeman1987cognitive, funder1995accuracy, krueger2007social, olsson2023cognitive}. Friends are fairly accurate in judging one another \cite{robbins2005social}, although there is a general positivity bias in close relationships \cite{gagne2004bias}. In online settings, individuals infer beliefs relying on stereotypes and ego projection \cite{goel2010real}. Perceptions involving out-groups are even more prone to biases \cite{bursztyn2022misperceptions, galesic2018sampling}.

\textbf{Ego Projection:} Individuals often appear to project their own thoughts, feelings, and attitudes about various issues onto others, assuming that others share similar beliefs \cite{Epley_2004, kruger1999lake}. This can be a Bayesian rational strategy in highly homophilous environments \cite{mcpherson2001birds, dawes1989statistical, dawes1996false}. Ego projection can contribute to false consensus \cite{ross1977false}. Conversely, a desire for uniqueness has been proposed to explain underestimation of agreement with others \cite{stern2014liberal}, although this false-uniqueness finding was not reproduced in a subsequent replication \cite{blanchar2021replication}. In PES, an inference that others' beliefs differ from one's own can be represented as negative ego projection; false uniqueness can also arise through social sampling without such a motivation \cite{galesic2018sampling}. Ego projection is represented by the directed link from personal beliefs to social beliefs: the latter represent inferences about what others personally believe, using their expressions as evidence and one's own personal belief as a reference \cite{Epley_2004}. 

\section*{Formal Implementation of the PES Framework}

In the formal PES model, each of the personal, expressed, and social beliefs is represented by a discrete random variable, whose spectrum (i.e., the set of all possible values that the variable can take) is indicated respectively with $S_b$, $S_e$, and $S_s$ (e.g., $S=\{-1,+1\}$ for a binary variable). Given an initial state, the beliefs are sequentially updated. The probabilistic rule driving the belief update process is a version of Glauber dynamics, where the local energy corresponds to the agent's potential dissonance $D$, multiplied by a factor $\beta \geq 0$ (inverse temperature), representing the agent's attention to the corresponding dissonance. We distinguish $\beta_{\text{pers}}$, $\beta_{\text{expr}}$, and $\beta_{\text{soc}}$ for personal, expressed, and social belief updates, respectively; $\beta$ denotes the corresponding parameter in the generic update rule below.

In each step of the updating process, the focal belief is randomly updated to one of the values of the related spectrum $S$. Each of the possible updates, indicated with $\tilde{y}\in S$, is sorted with probability:
\begin{equation}
    P(\tilde{y}) = \frac{e^{-\beta D(\tilde{y})}}{\sum_{x\in S }e^{-\beta D(x)}} 
\label{eq Glauber dynamics}
\end{equation}
where $D(x)$ is the potential dissonance calculated at the focal belief $x$, given the state of all the other beliefs (see \eqref{eq: dissonances}). Beliefs that lower the dissonance have a higher probability of being chosen in the process of update; if $\beta=\infty$, the focal belief is deterministically updated to the one that minimizes the dissonance function.

We denote with $\mathcal{N}_i$ the set of $i$'s neighbours. Letting $\mathbf{s_{(i)}}= ( s_{j(i)} )_{j\in \mathcal{N}_i}$ be the vector of $i$'s social beliefs (and $s_i$ its mean), and $\mathbf{b^*_i} $ be the vector of non-focal personal beliefs of agent $i$, the potential dissonance $D$ is the sum of the distinct socio-cognitive processes\footnote{They should be indicated, respectively, as $D_{personal}, D_{expressed}, D_{social}$: with a little abuse of notation, for the sake of readability, the subscripts have been omitted.}:

\begin{align}
    \underbrace{D(\tilde{b}_i)}_{\text{personal}} &=\ 
        \underbrace{d(\tilde{b}_i,\mathbf{s_{(i)}})}_{\text{social influence}} + 
         \underbrace{d(\tilde{b}_i,e_i)}_{\text{self-reinf.}} + 
         \underbrace{d(\tilde{b}_i,b_i)}_{\text{inertia}} + 
         \underbrace{d(\tilde{b}_i, \mathbf{b^*_i})}_{\text{context}}  \nonumber\\[1ex]
    \underbrace{D(\tilde{e}_i)}_{\text{expressed}} &=\ 
        \underbrace{d(\tilde{e}_i,b_i)}_{\text{authenticity}} +  
         \underbrace{d(\tilde{e}_i,\mathbf{s_{(i)}})}_{\text{conformity}}  \label{eq: dissonances} \\[1ex] 
    \underbrace{D(\tilde{s}_{j(i)})}_{\text{social}} &=\ 
        \underbrace{d(\tilde{s}_{j(i)},e_j)}_{\text{validity}} + 
         \underbrace{d(\tilde{s}_{j(i)},b_i)}_{\text{ego proj.}} \nonumber
\end{align}

When the vector $\mathbf{y}$ is a single-element vector, we consider as explicit expression of the contribution $d(\tilde{y},y) = \beta_{\text{process}} dist(\tilde{y},y)$, where $\beta_{\text{process}}$ reflects the weight of the dissonance stemming from that socio-cognitive process, and $dist(\tilde{y},y)$ measures the discrepancy between the candidate value $\tilde{y}$ of the belief being updated and the influencing belief value $y$ (the previous value of the same belief in the case of inertia). Generally, we use the squared Euclidean distance, $dist(\tilde{y},y)=(\tilde{y}-y)^2$, but other functions can be considered\footnote{Functions that reward the anti-alignment between beliefs can be used to represent anti-conformity or the desire for uniqueness, for example.}. In the case $\mathbf{y}$ has multiple entries, the dissonance related to the process is in general the sum of the dissonances of each pairwise interaction. \\
As an example, the dissonance of \eqref{eq: dissonances} driving the update of the expressed belief, with Euclidean distances reads
{\small
\begin{equation}
\underbrace{D(\tilde{e}_i)}_{\text{expressed}} = \underbrace{\beta_{auth}(\tilde{e}_i-b_i)^2}_{\text{authenticity}} + \underbrace{\beta_{\text{conf}} \sum_{j\in \mathcal{N}_i}(\tilde{e}_i-s_{j(i)})^2}_{\text{conformity}} 
\label{eq: pairwise general expression processes}
\end{equation}
}

\begin{table*}[t]
    \centering
    \caption{Reproducing belief (opinion) dynamics models within the PES framework.}
    \label{tab:all_models_unified}
    \scriptsize
    \setlength{\tabcolsep}{3pt}
    \begin{tabular}{|p{1.7cm}|p{2cm}|p{2cm}|p{8.2cm}|p{1.4cm}|}
    \hline
    \textbf{Model} & \textbf{Spectrum} & \textbf{Processes} & \textbf{Dissonances} & \textbf{Beta}\\
    \hline
    \multicolumn{5}{|c|}{\textbf{CLASS 1: No distinctions between belief types} ($s_{j(i)} = e_j = b_j$ at all times)}\\
    \hline
    \textit{All Class 1} & $S_e=S_b$ \newline $S_s=S_b$ & Full authenticity \newline No conformity \newline Perfect validity \newline No ego projection & {\tiny $D(\tilde{e}_i) = \underbrace{(\tilde{e}_i-b_i)^2}_{\text{authenticity}}$ \quad $D(\tilde{s}_{j(i)}) = \underbrace{(\tilde{s}_{j(i)}-e_j)^2}_{\text{validity}}$} & $\beta_{expr}=\infty$ \newline $\beta_{soc}=\infty$\\
    \hline
    \textbf{Voter} & $S_b=$$\{-1,+1\}$ & Social influence & {\tiny $D(\tilde{b}_i) = \underbrace{(\tilde{b}_i-b_j)^2}_{\text{social influence}}$} & $\beta_{pers}=\infty$\\
    \hline
    \textbf{Ising / Maj-vote / RFIM} & $S_b=$$\{-1,+1\}$ & Social influence \newline Context & {\tiny $D(\tilde{b}_i) = \underbrace{-\tilde{b}_i \sum_{j\in \mathcal{N}_i}b_j}_{\text{social influence}} \underbrace{-h_i\tilde{b}_i}_{\text{context}}$} & $\beta_{pers}$ finite \newline (Maj: $\infty$)\\
    \hline
    \textbf{DeGroot / Deffuant / HK / FJ} & $S_b=$$\{0,1,$$\ldots,n\}$ & Social influence \newline Context (FJ) & {\tiny $D(\tilde{b}_i) = \lambda_i\underbrace{\sum_{j\in \mathbb{S}_{i,\epsilon}} a_{ij}(\tilde{b}_i - b_j)^2}_{\text{social influence}} + (1-\lambda_i)\underbrace{(\tilde{b}_i-\bar{b}_i)^2}_{\text{context}}$} & $\beta_{pers}=\infty$\\
    \hline
    \hline
    \multicolumn{5}{|c|}{\textbf{CLASS 2: Distinct personal and expressed beliefs} ($s_{j(i)} = e_j$ at all times, $e_i \neq b_i$ allowed)}\\
    \hline
    \textit{All Class 2} & $S_s=S_e$ & Perfect validity \newline No ego projection & {\tiny $D(\tilde{s}_{j(i)}) = \underbrace{(\tilde{s}_{j(i)}-e_j)^2}_{\text{validity}}$} & $\beta_{soc}=\infty$\\
    \hline
    \textbf{Concealed Voter} & $S_b=S_e=$\newline$\{-1,1\}$ & Self-reinforcement \newline Authenticity \newline Conformity & {\tiny $D(\tilde{b}_i) = \underbrace{(\tilde{b}_i - e_i)^2}_{\text{self-reinforcement}}$ \quad $D(\tilde{e}_i) = \begin{cases} \underbrace{(\tilde{e}_i - b_i)^2}_{\text{authenticity}} \\ \underbrace{(\tilde{e}_i - e_j)^2}_{\text{conformity}} \end{cases}$} & $\beta_{pers}=\infty$ \newline $\beta_{expr}=\infty$\\
    \hline
    \textbf{2-layers Ising} & $S_b=S_e=$\newline$\{-1,1\}$ & Social infl. \newline Self-reinf. \newline Context \newline Authenticity \newline Conformity & {\tiny $D(\tilde{b}_i) = \beta_{\text{si}}\underbrace{(- \tilde{b}_i \sum_{j} e_j)}_{\text{soc. influence}} + \beta_{\text{sr}}\underbrace{(-  \tilde{b}_i e_i)}_{\text{self-r.}} \underbrace{- h_i\tilde{b}_i}_{\text{context}}$ \newline $D(\tilde{e}_i) = \beta_{\text{c}}\underbrace{(-\tilde{e}_i \sum_{j} e_j)}_{\text{conf.}} + \beta_{\text{a}}\underbrace{(-\tilde{e}_i b_i)}_{\text{auth.}}$} & $\beta_{pers}$ finite \newline $\beta_{expr}$ finite\\
    \hline
    \textbf{SOAM} & $S_b=S_e=$\newline$\{0,1,\ldots,n\}$ & Social infl. \newline Self-reinf. \newline Authenticity \newline Amplification & {\tiny $D(\tilde{b}_i) = \underbrace{\sum_{j\in \mathbb{S}_{i,\epsilon}} (\tilde{b}_i-e_j)^2}_{\text{social influence + self-reinforcement}}$ \newline $D(\tilde{e}_i) = \underbrace{(\tilde{e}_i-b_i)^2}_{\text{authenticity}} + \beta_{\text{ampl}}\underbrace{\gamma(\tilde{e}_i, b_i)}_{\text{amplification}}$, \; $\gamma(\tilde{e}_i, b_i) = \begin{cases} (\tilde{e}_i-n)^2 & b_i>n/2\\ 0 & b_i=n/2\\ (\tilde{e}_i)^2 & b_i<n/2 \end{cases}$} & $\beta_{pers}=\infty$ \newline $\beta_{expr}=\infty$\\
    \hline
    \textbf{EPO} & $S_b=S_e=$\newline$\{0,1,\ldots,n\}$ & Social infl. \newline Inertia \newline Context \newline Authenticity \newline Conformity & {\tiny $D(\tilde{b}_i) = \lambda_i\underbrace{\sum_{j \neq i} a_{ij}(\tilde{b}_i-e_j)^2}_{\text{soc. influence}} + \lambda_i\underbrace{a_{ii}(\tilde{b}_i-b_i)^2}_{\text{inertia}} + (1-\lambda_i)\underbrace{(\tilde{b}_i-\bar{b}_i)^2}_{\text{context}}$ \newline $D(\tilde{e}_i) = \phi_i\underbrace{(\tilde{e}_i - b_i)^2}_{\text{auth.}} + (1-\phi_i)\underbrace{\frac{1}{N} \sum_{j=1}^{N} (\tilde{e}_i-e_j)^2}_{\text{conformity}}$} & $\beta_{pers}=\infty$ \newline $\beta_{expr}=\infty$\\
    \hline
    \hline
    \multicolumn{5}{|c|}{\textbf{CLASS 3: Distinct personal and social beliefs} ($e_i = b_i$ at all times, $s_{j(i)} \neq e_j$ allowed)}\\
    \hline
    \textit{All Class 3} & $S_e=S_b$ & Full authenticity \newline No conformity & {\tiny $D(\tilde{e}_i) = \underbrace{(\tilde{e}_i-b_i)^2}_{\text{authenticity}}$} & $\beta_{expr}=\infty$\\
    \hline
    \textbf{NB theory} & $S_b = S_s =$$\{-1,$\newline$-.66,$$-.33,$$0,.33,$\newline$.66,1\}$ & Social influence \newline Context \newline Validity \newline Ego proj. & {\tiny $D(\tilde{b}_i) = \beta_{soc}^{NB}\underbrace{\left[-\tilde{b}_i\sum_{j} \rho_{ij} s_{j(i)}\right]}_{\text{social infl.}} + \beta_{pers}^{NB}\underbrace{\left[-\tau_i \tilde{b}_i - \tilde{b}_i \sum_{a} w_i^{(a)} b_i^{*(a)}\right]}_{\text{context}}$ \newline $D(\tilde{s}_{j(i)}) = \beta_{ext}^{NB}\underbrace{\left[-\alpha_{ij} \tilde{s}_{j(i)}b_j\right]}_{\text{validity}} + \beta_{soc}^{NB}\underbrace{\left[- \rho_{ij} \tilde{s}_{j(i)}b_i\right]}_{\text{ego proj.}}$} & $\beta_{pers}=1$ \newline $\beta_{soc}=1$\\
    \hline
    \end{tabular}
\end{table*}
    

\section*{Model Comparison within the PES Framework}

A key feature of the PES model is the ability to encompass a wide range of existing belief-dynamics models. By providing this conceptual overview, the framework facilitates principled model selection. We group the models into three distinct classes based on the constraints applied to the belief types (detailed in Table \ref{tab:all_models_unified} and in Appendix~\ref{sec: representing models}).

\subsubsection*{Class 1: Models without distinctions between belief types} 
These foundational models assume perfect alignment among all belief types at all times ($s_{j(i)} = e_j = b_j$). This constraint is enforced by setting infinite attention parameters ($\beta_{soc} = \beta_{expr} = \infty$). Within this constraint, the PES framework recovers the Voter model \cite{redner2019reality}, the Ising model and its variations (Random Fields Ising, Majority-vote) \cite{galam2002minority,galam1997rational,brandt2021evaluating,dalege2018attitudinal}, and discretized versions of continuous models including DeGroot, Friedkin-Johnsen, and bounded confidence models \cite{lorenz2007repeated,lorenz2007continuous}. These models represent dissonance reduction through social influence, but cannot capture psychological processes involving discrepancies between belief types. Some of the models of this class had already been represented in a dissonance-minimization framework \cite{groeber2014dissonance}.

\subsubsection*{Class 2: Models with distinct personal and expressed beliefs} 
This class allows divergence between personal and expressed beliefs while assuming perfect perception ($s_{j(i)} = e_j$, implemented by setting $\beta_{soc} = \infty$). It includes the Concealed Voter Model \cite{gastner2018consensus,gastner2019impact,garcia2020concealed}, the 2-layer Ising model \cite{jarema2022private,battiston2016interplay,ertacs2020dynamic,alberici2024ferromagnetic}, the Social Opinion Amplification Model (SOAM) \cite{lim2022opinion}, and the Expressed-Private Opinion (EPO) model \cite{anderson2019recent,ye2019influence}. These models capture distorted, strategic expression but assume unbiased perception.

\subsubsection*{Class 3: Models with distinct personal and social beliefs} 
This class permits misperception while requiring authentic expression ($e_i = b_i$, implemented by $\beta_{expr} = \infty$). It includes the Networks of Beliefs (NB) theory \cite{dalege2023networks}, which uses multiplicative distances and incorporates social influence, context effects, validity, and ego projection. This framework represents biased perception but not distorted expression.

\section*{Using PES Framework to elucidate socio-cognitive mechanisms}

Next, we show how PES can be used to compare explanations of the same social phenomena  by switching socio-cognitive processes on and off. Different processes can produce similar outcomes, and the outcome of one process can change when another is present. Comparing these possibilities within the same framework helps explain how the phenomenon arises and which interactions sustain or reduce it. These comparisons can then inform hypotheses for empirical tests and plan effective interventions. 

Here we use the example of \textit{pluralistic ignorance}. In its classic formulation, most members of a group privately reject a position while mistakenly believing that most others accept it, but it has become an umbrella term that encompasses broader forms of collective misperception, including errors about how common a belief is or how strongly others hold it \cite{miller2023century, prentice1993pluralistic}. Here we follow the approach that measures pluralistic ignorance as a difference between average perceived and actual beliefs about the same issue across people. Examples where people underestimate the prevalence of a popular point of view about issues ranges from climate mitigation policies to women's rights \cite{dixon2024complexity, bursztyn2022misperceptions, sparkman2022americans}. In its narrow sense, pluralistic ignorance involves only personal and social beliefs, but it can emerge from interaction with expressed beliefs \cite{gavrilets2026cultural} through the phenomena of spiral of silence \cite{noelle1974spiral} and preference falsification where there is social pressure to misrepresent private preferences \cite{kuran1995private}. Considering the joint dynamics of personal, expressed and social beliefs is precisely the regime the PES framework is built for.


Pluralistic ignorance can emerge from several different constellations of the underlying socio-cognitive mechanisms. These constellations correspond to specific constrained versions of PES discussed above. Pluralistic ignorance has typically been modeled using Class 2 models, whereby expressed beliefs diverge from personal ones but are themselves perfectly observed and hence equal to social beliefs; or by Class 3 models where perception of others' beliefs can be biased but expression is authentic so personal and expressed beliefs are the same. 
The recent norm-utility model of Gavrilets et al. \cite{gavrilets2026cultural} goes further and couples a personal norm, a second-order belief about the average personal norm in the population, and a ternary action, enabling formalization of several mechanisms involved in pluralistic ignorance such as cognitive dissonance, social influence, ego projection and validity (see Discussion and Appendix~\ref{app:gavrilets} for a detailed comparison and mapping between PES and Gavrilets et al.'s model). 

In what follows, we use the PES framework as a research instrument: we consider candidate mechanisms, formalize each in the simplest PES reduction that can express it, and ask whether it actually produces pluralistic ignorance, under which conditions, and how it compares to other plausible PES reductions. Two aggregate quantities are tracked throughout:
\begin{equation}
  \mathrm{PI}(t) = \bar s - \bar b,
  \qquad
  F(t) = \bar e   - \bar b  , \label{eq:def:PI_F}
\end{equation}
where $\mathrm{PI}$ is pluralistic ignorance and $F$ is aggregate preference falsification,
the gap between what is expressed and what is believed, at time $t$.

\subsection*{Setup}

We consider $N$ agents on a complete graph, with beliefs in the continuous spectrum $[-1,+1]$ unless otherwise specified, in the deterministic limit
$\beta_{pers,expr,soc} \to \infty$ of ~\eqref{eq Glauber dynamics} with
quadratic distances, so that each update is the dissonance-minimizing convex
combination of the influencing beliefs (see Appendix~\ref{app:PI}):

\begin{align}
  s_i &= (1-\varphi_i)\,\bar{e} \;+\; \varphi_i\, b_i , \label{eq:cs_s}\\
  e_i &= \alpha_i\, b_i \;+\; (1-\alpha_i)\, s_i , \label{eq:cs_e}\\
  b_i &= (1-\gamma_i)\big[\eta_i\, s_i + (1-\eta_i)\, e_i\big] \;+\; \gamma_i\, c^{+} .
  \label{eq:cs_b}
\end{align}
The four weights are the relative attentions of the corresponding
socio-cognitive processes: $\varphi = \beta_{\text{ego}}/(\beta_{\text{ego}}+\beta_{\text{val}})$
is ego projection against validity, $\alpha = \beta_{\text{auth}}/(\beta_{\text{auth}}+\beta_{\text{conf}})$
is authenticity against conformity, $\gamma$ is the weight of the belief context $c^+$
against the sum of social influence and self-reinforcement, and $\eta$ splits the
latter two. These weights are in general homogeneous across
individuals, in the following simulations. 

Agents are updated asynchronously: an agent is drawn at random and
updates first his personal and then his expressed belief; afterwards, all agents
update their social beliefs.\footnote{Refreshing every agent's social belief at
each step removes pluralistic ignorance that would arise simply because
perceptions have not yet caught up with what is currently expressed. Such a
delay is a simple yet plausible route to conservative lag and thus pluralistic ignorance when personal beliefs change rapidly. We remove it here
so that the mechanisms we do study cannot be confounded with it.}

Pluralistic ignorance is an out-of-equilibrium phenomenon. Excluding degenerate cases (see Appendix~\ref{app:PI}), the only fixed points of ~\eqref{eq:cs_s}--\eqref{eq:cs_b} are those in which the three belief types coincide, so that PI = F = 0; this holds also for heterogeneous parameters (proof in Appendix~\ref{app:PI}). This is in line with the model of Gavrilets et al. \cite{gavrilets2026cultural}, in which pluralistic ignorance emerges from lags between belief change and social belief updating and appears as a transient peak rather than as a property of equilibrium. Thus, we set up in a regime of belief change: we initialize personal beliefs opposed on average, $\langle b \rangle_0 = -0.5$, normally distributed across individuals with standard deviation $\sigma_b=0.05$, and switch on a context effect $c^{+}=+1$ representing an adjacent belief in the individual belief network that has changed or become salient (e.g., ``temperatures are rising'', if the focal belief is ``support to climate action"), or an external campaign. Under the baseline assumptions of full authenticity and validity ($\alpha=1$, $\varphi=0$), all three belief types move together to $+1$ and no misperception ever appears (Fig.~\ref{fig:gamma1results}a).\\
Averaging ~\eqref{eq:cs_s} and applying ~\eqref{eq:def:PI_F} yields (see Appendix~\ref{app:PI}), at any time, an exact decomposition of the aggregate misperception gap:
\begin{equation}
\mathrm{PI} = (1-\bar\varphi)\,F + \mathrm{Cov}(\varphi_i, b_i),
\label{eq:PIcov}
\end{equation}
where $\bar\varphi = \frac{1}{N}\sum_i \varphi_i$ and $\mathrm{Cov}(\varphi_i, b_i) = \frac{1}{N}\sum_i (\varphi_i - \bar\varphi)(b_i - \bar b)$. Pluralistic ignorance can thus arise only from preference falsification, of which ego projection removes a fraction $\bar\varphi$, or from a covariance between projection and personal belief; the simulations below explore each route.

\begin{figure*}[t!]
    \centering
    \includegraphics[scale=0.17]{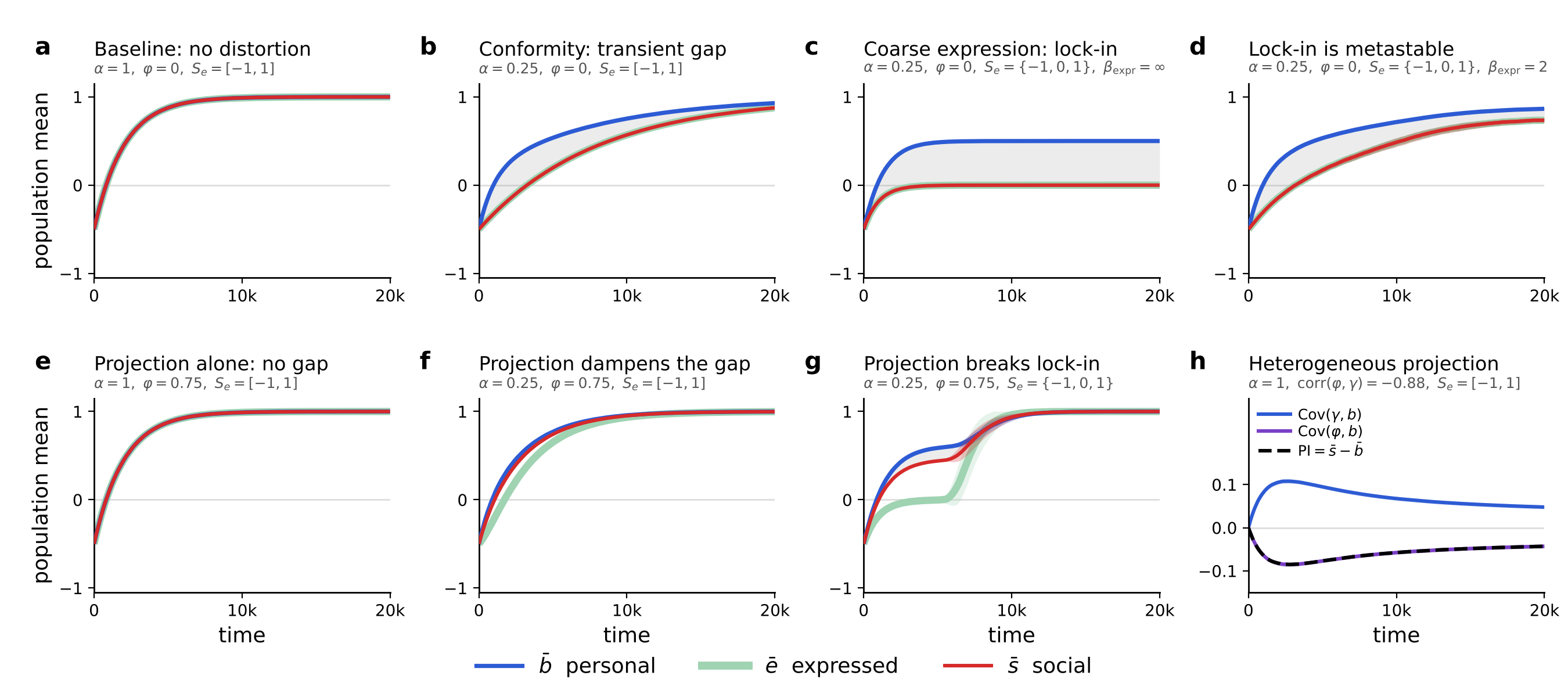}
    \caption{\textbf{Pluralistic Ignorance.} Simulations of panels \textbf{(a)--(g)} with homogeneous parameters $\gamma=\eta=0.5$, $c^+=1$; $S_b=S_s=[-1,1]$, $\beta_{pers}=\beta_{soc}=\infty$, $\beta_{expr}=\infty$ unless specified, $\varphi, \alpha$ as in the legends of the subplots. Average personal, expressed and social beliefs of a population of $N=1000$ individuals over $20N$ asynchronous timesteps, mean of $100$ runs (shading $=\pm 1$ SD across runs). Initial beliefs normally distributed with $\langle b \rangle_0=-0.5$, $\sigma_b=0.05$. The grey shaded area between $\bar{b}$ and $\bar{s}$ marks pluralistic ignorance, $\mathrm{PI}=\bar{s}-\bar{b}$. Panel \textbf{(h)}, heterogeneous parameters: $\varphi_i$ and $\gamma_i$ drawn from normal distributions with mean $0.5$ and variance $0.08$ and $0.12$ respectively, truncated to $[0,1]$, and coupled through a Gaussian copula with $\mathrm{corr}(\varphi,\gamma)=-0.88$; $\varphi$ and $\gamma$ are uncorrelated with $b_0$; $\eta=1$. Shown are $\mathrm{Cov}(\gamma,b)$, $\mathrm{Cov}(\varphi,b)$ and $\mathrm{PI}=\bar{s}-\bar{b}$; under authentic expression ($\alpha=1$) the last two coincide, as in Eq.~\eqref{eq:PI solo Cov}.}
    \label{fig:gamma1results}
\end{figure*}

\subsection*{Conformity alone produces only transient misperception}

Relaxing authenticity ($\alpha=0.25$, i.e. conformity weighted three times more
than authenticity) makes expression sluggish: agents whose personal belief has
already shifted keep expressing something close to the perceived norm. Because
social beliefs are read off expression, the perceived norm lags as well, and
$\mathrm{PI}$ and $F$ coincide and become strongly negative, down to $\approx -0.5$
(Fig.~\ref{fig:gamma1results}b). The gap, however, closes: as the personal
consensus saturates, expression eventually follows, and the misperception
disappears. Conformity by itself delays the social recognition of a belief change
but does not stabilize a false social reality.

\subsection*{Coarse-grained expression leads to a lock-in}

Personal beliefs are graded, but public expression is typically less nuanced
than the belief behind it \cite{urbig2003attitude}. In many settings one can only endorse, oppose, or stay
silent. The choice of silence is precisely what the spiral of silence
literature takes as the elementary public act \cite{noelle1974spiral,
kuran1995private}. We therefore keep the parameters above and coarsen only the
expression spectrum, $S_e = \{-1,0,+1\}$, leaving personal and social beliefs
continuous.

Under best response, average expression rises from $-0.5$, reaches silence
($\bar{e}=0$), and stops there permanently, while personal beliefs converge to
$+0.5$. PI saturates at $-0.5$ and never decays
(Fig.~\ref{fig:gamma1results}c). When the nearest available expression to the
dissonance-minimizing value is silence, everyone falls silent. This keeps the perceived norm at zero, which in turn keeps the
conformity term pulling expression back to silence. The population is trapped in
a state that no agent's personal belief supports.

The lock-in is metastable rather than absolutely stable. Reintroducing noise in
expression ($\beta_{\text{expr}}=2$) lets expression escape and drift upward, though
a substantial persistent gap remains (Fig.~\ref{fig:gamma1results}d). What matters
for the persistence of misperception is thus the granularity of the public
repertoire, not only the strength of conformity: coarsening expression lowers the
conformity threshold above which falsification becomes self-sustaining. This also
suggests that transient pluralistic ignorance reported in models with discretized action or opinion
spaces may be driven in part by the discretization itself rather than by the
substantive mechanism under study --- as may be the case in
\cite{gavrilets2026cultural}, where the action is ternary (behaving according to
the old norm, remaining silent, or behaving according to the new one).

\subsection*{Dissolution of PI through ego projection}

Ego projection can be intuitively thought as a cause of misperception, but the PES
framework shows that its aggregate effect is not always what that intuition suggests. First, when everyone has the same level of ego projection (or false consensus),  the individual errors it creates
cancel exactly in the aggregate, so pluralistic ignorance never appears. Specifically, relaxing perfect validity alone ($\varphi=0.75$, $\alpha=1$) leaves
$\mathrm{PI}=0$ at all times (Fig.~\ref{fig:gamma1results}e). Everyone perceives their social environment closer to their own position than it is, but the individual errors have opposite signs and cancel in the mean. In other words, false consensus at the individual level does not imply pluralistic ignorance at the
population level. The situation is analogous to network homophily, which biases what each
individual samples from their social circle but produces no aggregate misperception when it is symmetric across groups \cite{geiger2026sources, galesic2018sampling}. 

Second, when ego projection is combined with distorted expression, ego projection actually \emph{reduces} collective misperception, and can even dismantle a falsification-driven lock-in. Specifically,
combining conformity with ego projection ($\alpha=0.25$, $\varphi=0.75$) leads to a relatively
large preference falsification, $F \approx -0.25$, but dampens the misperception to a maximum of
$\mathrm{PI} \approx -0.11$, thus separating $\mathrm{PI}$ from $F$, as ~\eqref{eq:PIcov} prescribes
(Fig.~\ref{fig:gamma1results}f, compare with Fig.~\ref{fig:gamma1results}b). More strikingly, adding projection to the
coarsened-expression model dissolves the lock-in entirely
(Fig.~\ref{fig:gamma1results}g, compare with Fig.~\ref{fig:gamma1results}c). All three types of beliefs converge to $+1$ with
only a transient gap. Ego projection moves social beliefs ahead of observed expression,
which relaxes the conformity pull on expression. A perceptual bias can thus be corrective when the signal being perceived is itself systematically distorted.

\subsection*{Ego projection produces PI when it covaries with beliefs}
Ego projection leads to PI only when it co-varies with personal belief so that it is heterogeneous, with those who did not yet change their personal belief ego-projecting more than those who did. Under fully authentic expression, i.e. $\alpha=1$, $F=0$, ~\eqref{eq:PIcov} gives
\begin{equation}
    \mathrm{PI} = \mathrm{Cov}\big(b_i, \varphi_i\big)
\label{eq:PI solo Cov}
\end{equation}
This covariance need not be imposed exogenously; it can be generated by the dynamics.
We let projection be negatively correlated with context dependence
($\mathrm{corr}(\varphi,\gamma)=-0.88$), on the assumption that individuals whose
focal belief is well embedded in a coherent belief network (high $\gamma$) rely less on ego projection. Projection and initial beliefs remain uncorrelated at $t=0$.
As the context effect propagates, the highly context-dependent agents --- who are
also the weak ego projectors --- shift first, so $\mathrm{Cov}(b_i,\varphi_i)$ becomes
negative endogenously and pluralistic ignorance emerges, reaching $\approx-0.08$
before decaying (Fig.~\ref{fig:gamma1results}h). Underestimation of majority support
therefore arises from the asymmetry that those holding the old belief project
more, not from projection as such. This interpretation is consistent with reports of
stronger ego projection among conservatives \cite{blanchar2021replication} and of false
uniqueness among liberals \cite{stern2014liberal}. 

Appendix~\ref{app:PI} reports robustness checks against variations across runs and in the initial conditions, together with additional simulations.

\subsection*{Implications for measurement and interventions}
The modeling results open doors to posing theoretically-driven hypotheses that can be tested empirically. Each of the processes in the PES model can in principle be approximated empirically, either by direct questions or by inferring them from an existing data set. For example, people can be asked how important it is to them to express their true personal beliefs (authenticity, e.g. \cite{wood2008authenticity}), how much they assume others are like them (ego projection, e.g. \cite{andrews1993defense}), how much perceived others influence their personal beliefs (social influence, e.g. informational scale in \cite{bearden1989interpersonal}), or how important it is to them to appear similar to perceived others (conformity, e.g. \cite{mehrabian1995conformity}). Other parameters can be inferred from experimental or longitudinal observational data, including how much people's expressions and actions affect their personal beliefs over time (self-reinforcement, akin to the induced compliance literature, e.g. \cite{gerard1974compliance}), how accurate they typically are in perceiving others (validity, e.g. \cite{goel2010real}), and how much other beliefs 
affect the focal belief (context effects, e.g. because of cognitive dissonance \cite{Festinger_1957}). Many of these variables can also be inferred from cross-sectional data that includes measures of all three types of beliefs, although only on the group level. Once the parameters are estimated, the researchers can plug them in the model simulations above to elucidate the extent and the direction of misperceptions that can be expected.

Understanding the mechanisms underlying pluralistic ignorance in a specific situation matter for planning concrete interventions. Where misperception is sustained by a
coarse expressive repertoire, correcting beliefs about the norm is less effective
than restoring intermediate options for expressing them, since the lock-in is
reinstated as soon as public discourse re-polarizes into endorse/oppose/silent.
Where misperception is driven both by falsification and by ego projection, ~\eqref{eq:PIcov} implies that
reducing projection may not only leave the aggregate gap unchanged but also backfire, widening the misperception gap that ego projection had previously attenuated. Instead, what has to be targeted is the differential, that is, the group whose
projection is highest and whose beliefs are furthest from the emerging majority.

\section*{Discussion}

The PES meta-model contributes to belief dynamics research in two main ways. First, it provides a framework within which belief-dynamics models can be implemented and compared in a shared formal structure. This makes model comparison more principled and clarifies what is gained or lost when distinctions between personal, expressed and social beliefs are imposed or removed.

Second, the PES meta-model formalizes within a single framework socio-cognitive processes that have been studied extensively in the literature but are rarely formalized jointly in belief-dynamics models. Personal beliefs can be updated through social influence, self-reinforcement, context effects, and inertia. Expressed beliefs can be updated through authenticity and conformity. Social beliefs can be updated through validity and ego projection. Bringing these processes together makes it possible to study how they interact rather than examining them one at a time.

This is what we did to study pluralistic ignorance. In a regime of belief change, we found that: (i) conformity produces only transient pluralistic ignorance, by making expression viscous, that is, by holding public expression, and hence the perceived norm, close to the old belief after personal beliefs have already shifted; (ii) combined with conformity, coarse-grained expression (e.g., endorse, oppose, or stay silent) can trap the system in metastable states with persistently high pluralistic ignorance; (iii) contrary to intuition, in both of these settings ego projection does not amplify pluralistic ignorance but substantially dampens it, and can even dissolve the lock-in, while homogeneous ego projection alone produces none; (iv) ego projection generates pluralistic ignorance only when it is heterogeneous and covaries with personal belief, i.e., when those who have already adopted the new belief project less than those who still hold the old one, so that the average perception is biased towards the old belief. Such a covariance need not be imposed exogenously: it can emerge endogenously from the dynamics when ego projection is correlated with other cognitive parameters, for instance when weaker projectors are also more sensitive to the belief context. Simulating and quantitatively understanding these mechanisms is essential for designing interventions that mitigate pluralistic ignorance, for example to identify which group should be targeted and how.

Moreover, the flexible structure of PES allows it to reproduce another recent model of pluralistic ignorance, that of Gavrilets et al.~\cite{gavrilets2026cultural}, and thus to interpret and extend its results (see Appendix~\ref{app:gavrilets}). For example, within PES we could study the effect of discretizing the expressed-belief spectrum, which is fixed and ternary in~\cite{gavrilets2026cultural} ($S_e=\{-1,0,1\}$), and introduce noise through the stochastic update rule. More broadly, PES can be seen as a bridge between the literature on norm dynamics, exemplified by~\cite{gavrilets2026cultural}, and that on belief dynamics. In norm dynamics, the importance of second-order beliefs (\textit{normative expectations} in Bicchieri's terminology~\cite{bicchieri2016norms}) is well established, whereas belief-dynamics models typically assume that individuals' representations of others' beliefs coincide with what others express (or with their personal beliefs, when the expression layer is absent; but see~\cite{dalege2023networks}). To further enrich norm-utility models and, conversely, to couple belief-dynamics models with behavior, one could retain the three layers of PES, interpreting them respectively as personal normative beliefs, their expression, and normative expectations in Bicchieri's sense, and add a further layer of action, i.e., behavior that does or does not comply with the norm, with its material costs and benefits.

In our analyses we mostly assumed a complete network with pairwise interactions. We thus examined distortion and misperception at a purely cognitive level, rather than at the structural level addressed by other models such as the Social Sampling Model~\cite{galesic2012social, galesic2018sampling} or models of homophilous networks~\cite{lee2019homophily}. The framework thus can be extended to other network topologies, for instance to study the interplay between homophily and the cognitive processes of distortion and misperception \cite{landgren2026homophily}, as well as to higher-order interactions~\cite{battiston2021physics}. On this latter point, since individuals engage in successive interactions with different subsets of their contacts, and may in principle express (or withhold) different beliefs depending on the interaction, a natural extension would be to assign each individual not one but as many expressed beliefs as interaction contexts. Combined with nonlinear group interaction functions, this could yield new insights, for example on how the relative frequency of pairwise versus group interactions affects pluralistic ignorance.

Finally, we assumed cognitive parameters, i.e., the weights or attention individuals give to each process in belief updating, to be constant over time. Drawing on a recent line of work in norm dynamics that models the evolution of norm strength~\cite{szekely2025social}, extended versions could let these weights evolve dynamically, for example through social~\cite{zimmaro2026cultural} or reinforcement learning~\cite{schultner2025feature}.
 
Overall, PES offers a general framework for the joint dynamics of personal, expressed and social beliefs. It captures socio-cognitive processes that have been studied separately across disciplines and allows their combined effects to be investigated, while remaining firmly anchored to, and able to reproduce, a range of computational models from the sociophysics and opinion dynamics literature. In this sense, we hope that PES can help bring mathematical modeling and the social and behavioral sciences closer together in the quantitative study of social phenomena.

\section*{Acknowledgments}
This work was supported by grants from the National Science Foundation DRMS 1949432; National Science Foundation MMS 2019982; ERC Advanced Grant COLLADAPT 101140741; the  Austrian  Research  Promotion  Agency FFG grant 873927; and the OPUS Weave project funded by Austrian Science Fund 10.55776/PIN2028524 and Polish National Science Centre 2023/51/I/HS6/02269z.

\section*{Author contributions}
F.Z. and H.O. designed research; F.Z. and H.O. performed research; F.Z., M.G. and H.O. wrote the paper.

{\small

}

\onecolumn
\appendix
\renewcommand{\thesection}{\Alph{section}}
\numberwithin{equation}{section}
\numberwithin{figure}{section}
\numberwithin{table}{section}
\renewcommand{\theequation}{\thesection.\arabic{equation}}
\renewcommand{\thefigure}{\thesection.\arabic{figure}}
\renewcommand{\thetable}{\thesection.\arabic{table}}

\begin{center}
{\Large\bfseries Appendix}
\end{center}

\section{Representing belief dynamics models within the PES model} \label{sec: representing models}

We illustrate how several prominent belief (opinion) dynamics models can be represented within the PES meta-model framework. We do not assume any particular network structure and selection algorithm, and we focus only on belief updating in the interaction process. While most models are mapped to the PES model in its original form, some require additional integrations (specifically, for the Bounded Confidence (BCM) and the Social Opinion Amplification (SOAM) models). In what follows, we will use this further notation: $\mathbf{b_i}= ( b_j )_{j\in \mathcal{N}_i}$ vector of personal beliefs of agent i's neighbours, $\mathbf{e_i}= (e_j )_{j\in \mathcal{N}_i}$ vector of expressed beliefs of agent i's neighbours.

\subsection{Models without distinctions between different types of beliefs}
\begin{figure}[H]
    \centering
    \includegraphics[width=0.8\textwidth]{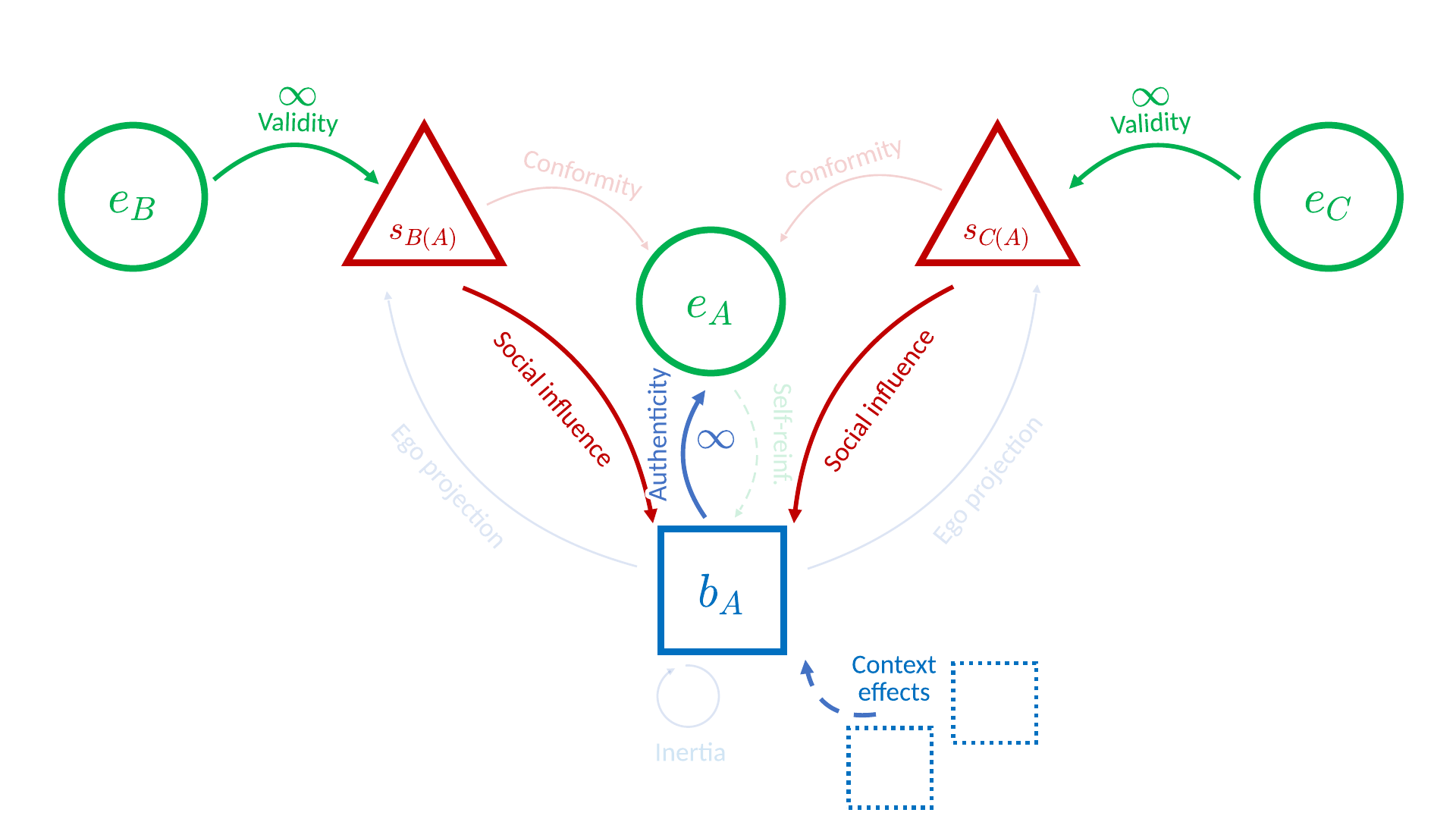}
    
    \vspace{0.5cm}
    
    \scriptsize
    \begin{tabular}{|p{1.5cm}|p{1.7cm}|p{2cm}|p{5cm}|p{1.6cm}|}
    \hline
    \textbf{Model}  & \textbf{Spectrum} & \textbf{Processes} & \textbf{Dissonances} & \textbf{Beta}\\
    \hline
    \textit{All of these models} & $S_e$=$S_b$ \newline $S_s$=$S_b$  & Full authenticity \newline No conformity \newline \newline Perfect validity \newline No ego projection & $D(\tilde{e}_i) =\underbrace{(\tilde{e}_i-b_i)^2}_{\text{authenticity}}$ \newline \newline $D(\tilde{s}_{j(i)}) = \underbrace{(\tilde{s}_{j(i)}-e_j)^2}_{\text{validity}}$ & $\beta_{expr}$=$\infty$ \newline \newline \newline $\beta_{soc}$=$\infty$ \\
    &&&&\\
    \hline
    \textit{\textbf{Voter}} & $S_b$=$\{-1,+1\}$ & Social influence& $D(\tilde{b}_i) = \underbrace{(\tilde{b}_i-b_j)^2}_{\text{social influence}}$ & $\beta_{pers}$=$\infty$\\
    &&&&\\
    \hline
    \textit{\textbf{Ising \newline (Random fields, \newline Majority vote)}} & $S_b$=$\{-1,+1\}$ & Social influence \newline Context & $D(\tilde{b}_i) = \underbrace{-\tilde{b}_i \sum_{j\in \partial i}b_j}_{\text{social influence}} \underbrace{-h_i\tilde{b}_i}_{\text{context}}$ & $\beta_{pers}$ finite, \newline $\beta_{pers}$=$\infty$ (Maj. vote)\\
    &&&&\\
    \hline
    \textit{\textbf{Continuous \newline (DeGroot, \newline Deffuant, HK, FJ)}} & $S_b $=$ \{0,1,2,$\newline $3,\ldots,n\}$ & Social influence \newline Context (FJ) & $D(\tilde{b}_i) = \lambda_i\underbrace{\sum_{j\in \mathbb{S}_{i,\epsilon}} a_{ij}(\tilde{b}_i - b_j)^2}_{\text{social influence}} +\newline +(1-\lambda_i)\underbrace{(\tilde{b}_i-\bar{b}_i)^2}_{\text{context}}$ & $\beta_{pers}$=$\infty$\\
    \hline
    \end{tabular}
    
    \caption{\textbf{Models without distinctions between different types of beliefs} (\textit{table}) and \textbf{corresponding version of the PES model} (\textit{figure}). In the figure, the social-psychological processes that are considered within these models are highlighted with thick arrows, while the ones of the general PES model that are not accounted are more transparent. The dashed lines indicate processes that are present in some models and absent in others (here, it is context effects). This holds also for the following figures related to the other classes of models.}
    \label{fig:PES no distinction}
\end{figure}

Most of the fundamental models proposed and studied in the literature do not allow any possible misalignment among personal, expressed and social beliefs. In other words, they assume that the perceived beliefs of others and their expressed beliefs perfectly coincide with the individuals' personal beliefs. This means that
\begin{equation}
    \begin{cases}
        s_{j(i)} =   b_j\\
        e_i =   b_i
    \end{cases}
    \quad \quad \quad \quad \quad \mbox{at any time}, \;\;\;\forall\;i,j
\label{eq model with no distinctions among beliefs}
\end{equation}
which can be realized within the PES model by setting
\begin{align}
    \notag
    &D(\tilde{s}_{j(i)}) = \underbrace{d(\tilde{s}_{j(i)},e_j)}_{\text{validity}} = (\tilde{s}_{j(i)}-e_j)^2\\ \label{eq no distinction}
    &D(\tilde{e}_i) = \underbrace{d(\tilde{e}_i,b_i)}_{\text{authenticity}} = (\tilde{e}_i-b_i)^2\\ \notag
    &\beta_{soc}=\beta_{expr}=\infty 
\end{align}
so that the social and expressed belief are every time deterministically updated to match respectively the neighbours' expressed beliefs and the focal individual's personal belief. The contributions of ego projection, self-reinforcement and conformity are not present: the effects of these social-psychological processes cannot be captured by the models with no distinctions between different types of beliefs. Figure \ref{fig:PES no distinction} and the accompanying table show the PES model in this setting, and highlight the social-psychological processes involved. Within this class of models, we manage to reproduce within the PES framework the \textit{Voter model}, the \textit{Ising model} and its variations (\textit{Majority-vote}, \textit{Random Fields Ising}), the discretized versions of a class of continuous models that comprises the \textit{DeGroot} and \textit{Friedkin-Johnsen (FJ)} models, and the \textit{bounded confidence models} of \textit{Deffuant} and \textit{Hegselmann-Krause (HK)}.\\
\\
\textbf{Voter models}\\
In the Voter model \cite{redner2019reality}, the belief is binary $\{-1,+1\}$ and the focal agent simply takes the belief of the agent he is interacting with. We recover the dynamics of the Voter model, considering the focal agent $i$ interacting with an agent $j$, by using \eqref{eq no distinction} and setting
\begin{align}
    \notag &S_b = \{-1,+1\} \quad (=S_s=S_e)&\\
    &D(\tilde{b}_i) = \underbrace{d(\tilde{b}_i,s_{j(i)})}_{\text{social influence}} = d(\tilde{b}_i,b_j) = (\tilde{b}_i - b_j)^2 \\
    \notag &\beta_{pers}= \infty
\end{align}
Notice that the contributions of the inertia and the context are set to zero. However, the context could be included under the form of fixed preferences attached to the agents \cite{zimmaro2023voter}. Through the framework of the PES model, the Voter Model can be interpreted as a process of optimal dissonance reduction, where the dissonance is given by the squared Euclidean distance with respect to the belief of the interacting neighbour, with completely faithful belief expressions and perfect representations. Notice that we chose the squared Euclidean distance for mathematical convenience and homogeneity with the other models, but, at least in this case, we could have as well used the absolute difference or a multiplicative distance.\\
\\
\textbf{Ising models: Random-Fields, Majority-vote}\\
In Ising models evolving with Glauber dynamics, beliefs are represented by binary spins. In these models, thepotential dissonance is analogous to the Ising Hamiltonian with pairwise couplings:
\begin{align}
    &S_b =\{-1,+1\} \quad (=S_s=S_e)  \\ 
    &D(\tilde{b}_i) = \underbrace{d(\tilde{b}_i,\mathbf{s_{(i)}})}_{\text{social influence}} = d(\tilde{b}_i,\mathbf{b_i}) =\sum_{j\in \mathcal{N}_i} (\tilde{b}_i - b_j)^2 \propto -\tilde{b}_i \sum_{j\in \mathcal{N}_i} b_j 
    \label{eq dissonance Ising standard} 
\end{align}
The proportionality coefficient in \eqref{eq dissonance Ising standard} is not influential, as it can be incorporated in $\beta_{pers}$. Being a model with stochastic update, beliefs that do not minimize the potential dissonance are also plausible and $\beta_{pers}$ can be finite. \\
The basic version of the Ising model can be enriched, for example by considering a fixed preference towards one or the other belief, attached to each individual \cite{zimmaro2024asymmetric}. Such preference enters in the Hamiltonian under the form of an external field $h_i$. Thus, we would consider 
\begin{equation} 
    D(\tilde{b}_i) =  \underbrace{-\tilde{b}_i \sum_{j\in \mathcal{N}_i} b_j}_{\text{social influence}}  \underbrace{ -h_i \tilde{b}_i}_{\text{context}}
\end{equation}

\noindent
Depending on the assignement of the fields, these models are versions of the so-called Random Field Ising Model (RFIM). 

\noindent
In the majority vote model \cite{galam2002minority}, beliefs are binary variables $\{- 1,+1\}$ and individuals in a population update their beliefs based on the majority belief within their local neighborhood. The update is thus deterministic. It is easy to see that the majority vote model corresponds to an Ising model with $\beta_{pers}= + \infty$. If we include fixed preferences in the majority vote model through the external fields $(h_i)_{i=1,\ldots,N}$, we would have the RFIM at zero temperature, another popular model in belief dyamics \cite{galam1997rational}.

\noindent
Viewed through the PES model, the Ising models can be interpreted as processes of generally imperfect dissonance reduction (stochastic update). Here, dissonance arises from the misalignment with the beliefs of all the focal agent's neighbors and is quantified by the squared Euclidean distance between beliefs, or equivalently (for binary beliefs), by the multiplicative distance. In the field-included version, dissonance also partly results from misalignment with a fixed belief, which may stem from convictions about other related issues. The majority-vote model represents an optimal dissonance reduction. Additionally, as with the other models in this subsection, the Ising models assume fully faithful belief expressions and perfect representations.\\
\\
\textbf{Continuous models: DeGroot and Friedkin-Johnsen models, Bounded Confidence models of Deffuant and Hegselmann-Krause}\\
Considering continous beliefs $x_i\in [0,1]$, $i=1,\ldots,N$, the update for the DeGroot model reads\footnote{For clarity, the notation $x\leftarrow y$ indicates that the focal belief $x$ is updated to take the value of $y$.}
\begin{equation}
    x_i \;\; \leftarrow \;\; \sum_{j=1}^N a_{ij}x_j
\label{DeGroot equation}
\end{equation}
where $\sum_{j=1}^N a_{ij} =1$ and $A=(a_{ij})_{i,j=1,\ldots,N}$ is the social network adiacency matrix that accounts for the strength of the influence between couples of individuals. The Deffuant model considers continous beliefs and pairwise interactions: if the distance between the beliefs of the interacting agents is below a certain threshold $\epsilon\in (0,1]$, then after the interaction the agents converge to the average belief between the two initial ones; else, if the distance exceeeds the threshold, nothing happens. The principle of the Hegselmann-Krause model is the same as the Deffuant model, with the difference that the focal agent simultaneously interacts with all his neighbours whose belief falls within his confidence bound $\epsilon$.

\noindent
Since in the PES model we take discrete beliefs, we consider a discretized analogue of such continuous models: taking $n\in \mathbb{N}$, we consider
\begin{equation}
    S_b = \{0,1,2,3,\ldots,n\} \quad \quad  (=S_s=S_e)
\end{equation}
In the discretized version, the resulting belief, calculated according to the model's prescriptions, is rounded to the one of the two close integer values. For $n$ sufficiently large, we expect that the discretized analogous behaves like the original versions of the models; for the Deffuant and HK models, this has been studied in \cite{lorenz2007repeated, lorenz2007continuous}. However, the use of a discretized versions of these model is wide, see e.g., \cite{jacobmeier2005multidimensional,stauffer2004discretized}.

\noindent
We recover the dynamics of the discretized version of these continous models, considering the focal agent $i$, by using \eqref{eq no distinction} and setting
\begin{align}
    &D(\tilde{b}_i) = \underbrace{d(\tilde{b}_i,b_i,\mathbf{s_{(i)}})}_{\text{social influence}} = d(\tilde{b}_i,b_i,\mathbf{b_i}) =\sum_{j\in \mathbb{S}_{i,\epsilon}} a_{ij}(\tilde{b}_i-b_j)^2 \label{contnous model social personal dissonance}\\        
    &\beta_{pers}= \infty& \label{continous models infinite beta}
\end{align}
where the set of interacting agents read
\begin{equation}
    \mathbb{S}_{i,\epsilon} = \{j\in\mathcal{N}_i^*: |b_j-b_i|<\epsilon n\} \; \bigcup \; \{i\} 
\end{equation}
The confidence bound $\epsilon$, the set of interacting neighbours $\mathcal{N}_i^*$, and the weights $\{a_{ij}\}_{i,j=1,...,N}$ determine the type of model. Specifically,
\begin{equation}
    \begin{cases}
        \epsilon = 1, \; \; \mathcal{N}_i^*=\mathcal{N}_i  \quad \quad \mbox{DeGroot model}\\
        \epsilon < 1, \; \; \mathcal{N}_i^*=\mathcal{N}_i,  \;\; a_{ij}=a_{ik} \;\forall \;i,j,k  \quad \quad \mbox{HK model}\\
        \epsilon < 1, \; \; \mathcal{N}_i^*=\{ j \sim U(\mathcal{N}_i)\}, \; \; a_{ij}=a_{ik} \;\forall \;i,j,k\quad \quad \mbox{Deffuant model}
    \end{cases}
\end{equation}
Notice that, to account for the bounded confidence effect, we need to add a further dependency of the social influence function \eqref{contnous model social personal dissonance} on the current belief of the focal agent $b_i$. We indicated with $j \sim U(\mathcal{N}_i)$ the uniformly at random extraction of a neighbour $j$ among the elements of the set $\mathcal{N}_i$. The fact that $\mathcal{N}_i^*$ is composed by a single randomly chosen agent or by the entire set of neighbours distinguishes the Deffuant form the Hegselmann-Krause model. The constraint $a_{ij}=a_{ik} \;\forall \;i,j,k$ can be easily relaxed for a more generalized version of the bounded confidence models.

\noindent
The validity of the form of the social influence in \eqref{contnous model social personal dissonance} follows from the following arguments. Consider \eqref{contnous model social personal dissonance} on a continous support, written as $D(z)$ for $z \in \mathbb{R}$. $D(z)$ is convex, $\lim_{z\rightarrow \pm \infty}D(z) = +\infty$ and thus it has a unique minimum, which reads
\begin{equation}
    z^* = \arg\min_{z\in \mathbb{R}} D(z) =
        \langle b \rangle_{\mathbb{S}_{i,\epsilon}, \{a_{ij}\}} 
\end{equation}
where $\langle b \rangle_{\mathbb{S}_{i,\epsilon}, \{a_{ij}\}} $ stands for the weighted average of the beliefs on the set $\mathbb{S}_{i,\epsilon}$. On the discrete support $S_b$, $D(\tilde{b}_i)$ is minimized by
\begin{equation}
    \tilde{b}^*_i = \arg\min_{\tilde{b}_i\in S_b} D(\tilde{b}_i) = \arg\min_{\tilde{b}_i=[z^*],[z^*]+1} D(\tilde{b}_i)
\label{discrete minimum}
\end{equation}
$[\cdot]$ denoting the floor function\footnote{Notice that it can happen that, while $z^*$ is always unique, $D(\tilde{b}_i)$ has two minima in the discrete support if $D([z^*])= D([z^*]+1)$. This is a technical problem happening in very special cases: we can overcome it by choosing by convention, between $[z^*]$ and $[z^*]+1$, the one closer to the center $n/2$ ($n$ even). One could interpret that as the introduction of a small pertubation of the dissonance function at $[z^*]$ that slightly favours more moderate beliefs.}.

\noindent
Thus, within the PES model framework, the DeGroot, Deffuant, and Hegselmann-Krause models can be rephrased as processes of optimal dissonance reduction, where the dissonance experienced by the focal agent is measured by the dissimilarity (mathematically represented as the squared Euclidean distance) between their own initial belief and that of the interacting agent(s). These models assume fully faithful belief expressions and perfect representations. The primary difference between the models, aside from the weighting schemes, lies in the set of interacting agents: DeGroot considers all neighbors, Hegselmann-Krause allows for simultaneous interactions with neighbors whose beliefs are within a certain range of the agent's initial belief, and Deffuant involves pairwise interactions with one agent selected from the set considered by the Hegselmann-Krause model. \\
\\
It is easy to map the PES also to some variations of the models above: one example is the Friedkin-Johnsen (FJ), which perturbs the DeGroot update \eqref{DeGroot equation} by adding fixed individual preferences. The FJ model's update reads
\begin{equation}
    x_i \;\;\leftarrow  \;\; \lambda_i \sum_{j=1}^N a_{ij}x_j + (1-\lambda_i)\bar{x}_i
\label{FJ dynamics PES}
\end{equation}
where $\lambda_i\in [0,1]$, $\bar{x}_i\in [0,1]$. $1-\lambda_i$ represents the individual's stubborness towards the fixed belief $\bar{x}_i$. The DeGroot model is recovered for $\lambda_i=1\;\forall \;i$. The discretized version of such model is recovered from the PES one by considering
\begin{equation}
    D(\tilde{b}_i) = \underbrace{\lambda_i d(\tilde{b}_i,b_i,\mathbf{s_{(i)}})}_{\text{social influence}} + \underbrace{(1-\lambda_i)(\tilde{b}_i-\bar{b}_i)^2}_{\text{context}}\\
\end{equation}
with \eqref{contnous model social personal dissonance} for $d(\tilde{b}_i,b_i,\mathbf{s_{(i)}})$, as in the DeGroot model, $\bar{b}_i = round(n\bar{x}_i)$ and \eqref{continous models infinite beta}. One can indeed check that such dissonance minimization leads to the discretized analogous of the dynamics \eqref{FJ dynamics PES}. Interestingly, the FJ model can be interpreted through the lenses of the PES model as a dissonance reduction (still under the shape of the squared Euclidean distance) with respect to the beliefs of the neighbouring agents but also to a fixed belief potentially coming from the convinctions about other issues. The influence of the context is regulated by $\lambda_i$.

\subsection{Models with distinct personal and expressed beliefs}
\begin{figure}[H]
    \centering
    \includegraphics[width=0.8\textwidth]{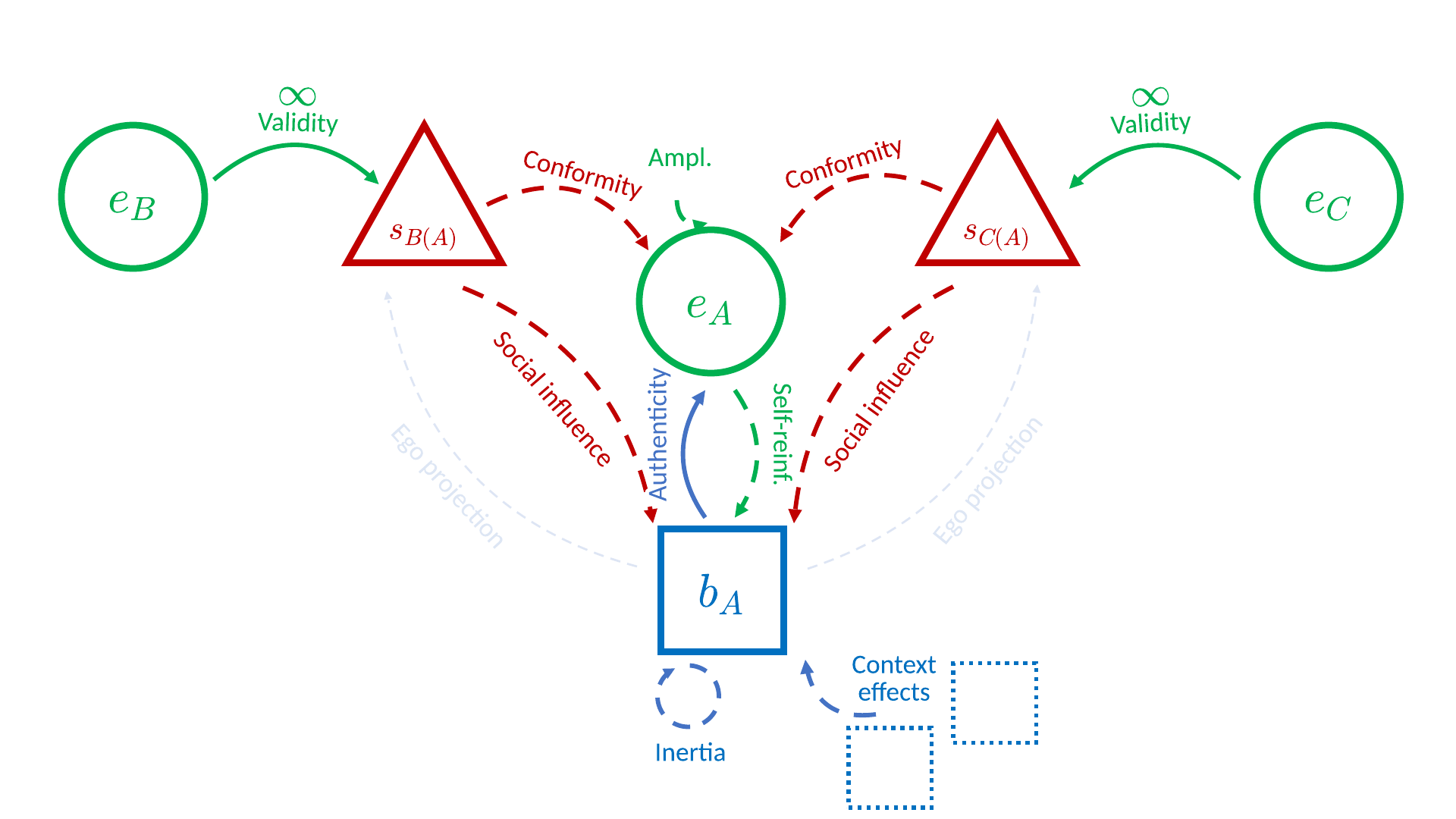}
    \caption{\textbf{Version of the PES model corresponding to models with distinct personal and expressed beliefs} (see Table~\ref{tab: models with distinct personal/ expressed}).}
    \label{fig:PES_model exressed and personal}
\end{figure}

There are several belief dynamics models that distinguish personal beliefs (also called \textit{private}, or \textit{internal}) from expressed (or \textit{public}) beliefs \cite{allport1937toward,schanck1932study,kuran1995private,asch2016effects}. They typically consist of extensions and adaptations of some of the fundamental models considered in the previous section. These models all implicitly assume that the social perception always coincides with the expressed belief of the corresponding agents, i.e.,
\begin{equation}
    s_{j(i)} =  e_j  \quad \quad \quad \mbox{at any time }, \;\;\; \forall \;i,j
\end{equation}
which can be realized within the PES model by setting
\begin{align}
    &D(\tilde{s}_{j(i)}) = \underbrace{d(\tilde{s}_{j(i)},e_j)}_{\text{validity}} = (\tilde{s}_{j(i)}-e_j)^2\\
    &\beta_{soc}=\infty \label{eq distinct personal and expressed}
\end{align}
In this realization, there is thus no effect of biased perception, e.g. through ego projection. Figure \ref{fig:PES_model exressed and personal} and the accompanying table show the PES model in this setting, along with the different social-psychological processes. Within this class of models, we can reproduce within the PES sheme the \textit{Concealed Voter Model}, a generalization of the Voter model with expressed beliefs, the \textit{2-layers Ising model}, a multilayer generalization of the Ising model, and the \textit{Social Opinion Amplification Model (SOAM)} and \textit{Expressed-Private Opinion (EPO)}, generalizations of continous bounded confidence models.\\
\\
\textbf{Concealed Voter Model}\\
In the Concealed Voter Model (CVM) \cite{gastner2018consensus,gastner2019impact,garcia2020concealed}, each agent has a personal and an expressed binary belief,
\begin{equation}
    S_b = S_e = \{-1,+1\},
\end{equation}
and can copy, internalize or externalize. In the process of \textit{copying}, agents update their expressed belief by copying the one of the interacting agent. In the process of \textit{internalization}, they update their personal belief by copying his expressed one. Vice versa, in the process of \textit{externalization}, the expressed belief is updated by copying his personal one. The three processes can occur at different rates, according to the model's prescriptions. As the basic Voter Model, the CVM prescribes a pairwise encounter, thus the dissonances of the general model, together with \eqref{eq distinct personal and expressed}, are
\begin{align}
    &D(\tilde{b}_i) = \underbrace{d(\tilde{b}_i,e_i)}_{\text{self-reinforcement}}=  (\tilde{b}_i - e_i)^2\\
    &\beta_{pers}= \infty
\end{align}
governing the internalization process, 
\begin{align}
    &D(\tilde{e}_i) = \begin{cases}
         \underbrace{d(\tilde{e}_i,b_i)}_{\text{authenticity}} =  (\tilde{e}_i - b_i)^2 \\
        \underbrace{d(\tilde{e}_i,s_{j(i)})}_{\text{conformity}} = d(\tilde{e}_i,e_j) = (\tilde{e}_i - e_j)^2
    \end{cases} \\
    &\beta_{expr}= \infty
\end{align}
\noindent
governing the externalization or copying mechanisms, depending on the action the focal agent is chosen to make.

\noindent
Through the lens of the PES model, the Concealed Voter Model can be interpreted as a process of optimal dissonance reduction for the belief expression and the construction of the personal belief. In the expression, the dissonance can be given alternatively by the dissimilarity (squared Euclidean distance) with respect to the interacting agent's expression or with respect to the focal agent's personal belief. In the personal belief definition, the dissonance is given by the dissimilarity with respect to the agent's expressed belief. The model assumes perfect representations, given the belief expressions.\\
\\
\textbf{2-Layers Ising models}\\
The 2-layers Ising model can represent the mechanisms of belief formation process with distinct expressed (first layer) and personal (second layer) beliefs \cite{jarema2022private,battiston2016interplay,ertacs2020dynamic, alberici2024ferromagnetic}.
Both the spectra for personal and expressed beliefs are the same
\begin{equation}
    S_b = S_e = \{-1,+1\}
\end{equation}
The update of individual $i$'s expressed belief is stochastic and it is guided both by the influence of the neighbours' expressed beliefs (conformity) and his personal current belief (authenticity): together with \eqref{eq distinct personal and expressed}, 
\begin{equation}
    D(\tilde{e}_i) = \underbrace{d(\tilde{e}_i,\mathbf{s_{(i)}})}_{\text{conformity}} + \underbrace{d(\tilde{e}_i,b_i)}_{\text{authenticity}} = d(\tilde{e}_i,\mathbf{e_{i}}) + d(\tilde{e}_i,b_i) = -\beta_{\text{conf}}\tilde{e}_i \sum_{j\in \mathcal{N}_i} e_j -\beta_{\text{auth}}\tilde{e}_i b_i\\
\end{equation}
with $\beta_{\text{conf}}, \beta_{\text{auth}}$ and the common factor $\beta_{expr}$, in general, finite. Notice that in this model we first had to specify the processes' weights ($\beta_{\text{process}}$, as in eq. \eqref{eq: pairwise general expression processes}, which were implicitely set to 1 in the other models so far considered), as there are multiple processes contributing to the same dissonance with potentially different weights.\\
The update of the personal belief is guided by both the social influence of the neighbours' expressed beliefs and from the internalization effect, as well as by the potential effect of a fixed preference (context), each with the corresponding weight $\beta_{\text{process}}$ ($h_i$ for the preference):
\begin{equation}
    D(\tilde{b}_i) = \underbrace{d(\tilde{b}_i,\mathbf{s_{(i)}})}_{\text{social influence}} + \underbrace{d(\tilde{b}_i,e_i)}_{\text{self-reinforcement}} + \underbrace{d(\tilde{b}_i)}_{\text{context}} 
\end{equation}
where
\begin{align}    
    &\underbrace{d(\tilde{b}_i,\mathbf{s_{(i)}})}_{\text{social influence}}= d(\tilde{b}_i,\mathbf{e_i})= \sum_{j\in \mathcal{N}_i} d(\tilde{b}_i,e_j) =  - \beta_{\text{socinf}}\tilde{b}_i \sum_{j\in \mathcal{N}_i} e_j\\
    &\underbrace{d(\tilde{b}_i,e_i)}_{\text{self-reinforcement}}= - \beta_{\text{selfreinf}} \tilde{b}_i e_i\\
    &\underbrace{d(\tilde{b}_i)}_{\text{context}} = - h_i\tilde{b}_i    
\end{align}
and $\beta_{pers}$, in general finite.

\noindent
Through the lens of the PES model, the 2-layers Ising models with private and public beliefs can be interpreted as a process of imperfect dissonance reduction for the belief expression and the construction of the personal belief. In the expression, the dissonance is given jointly by the dissimilarity (multiplicative or, equivalently for binary beliefs, squared Euclidean distance) with respect to the interacting agents' expressed beliefs (conformity) and with respect to the focal agent's personal belief (authenticity). In the personal belief definition, the dissonance is given by the dissimilarity with respect to the focal agent's expressed belief (self-reinforcement), the dissimilarity with respect to the interacting agents' expressed beliefs (social influence) and by a fixed preference towards one or the other belief (context). The model assumes perfect representations, given the belief expressions. The extension to an analogous version of the majority vote model is straightforward.\\
\\
\textbf{Continuous models: Social Opinion Amplification Model (SOAM) and Expressed-Private Opinion (EPO)}\\
The SOAM \cite{lim2022opinion} and EPO \cite{anderson2019recent,ye2019influence} models are respectively an adaptation of the Hegselmann-Krause and Friedkin-Johnsen models with the inclusion of different private (personal) and expressed beliefs.  In order to reproduce them within the PES model, we consider discrete spectra
\begin{equation}
    S_b = S_e = \{0,1,2,3,\ldots,n\} \quad \quad  (=S_s)
\end{equation}
In the SOAM, the individual updates his personal belief by taking the average of the expressed beliefs of the neighbouring agents whose expressed belief falls within a confidence threshold. 

\noindent
The update of the expressed belief, instead, is governed by two processes: the aim to authentically express the personal belief and an incentive to express an amplified, i.e., more extreme, belief. 
Such amplification can be caused, when expressing beliefs on online social platform, by the aim of attracting attention: the more extreme a post, the more popular it is \cite{tucker2018social, bail2022breaking}. 
In the original SOAM model, the expressed belief is deterministically derived by amplifying the personal belief with a constant value, making it more extreme.
Such rigid update is hard to reproduce within the PES model\footnote{It is actually hard to justify the independence of the amplification term and the agent's current belief.}, so we reproduce the incentive for belief amplification by adding a further term $\gamma(\tilde{e}_i, b_i)$, whose weight is regulated by $\beta_{\text{ampl}}$ and whose form differs from the standard one \eqref{eq: pairwise general expression processes}, to the expressive dissonance:
\begin{align*}
    &D(\tilde{b}_i) = \underbrace{d(\tilde{b}_i,\mathbf{s_{(i)}})}_{\text{social influence}} + \underbrace{d(\tilde{b}_i,e_i)}_{\text{self-reinforcement}}= d(\tilde{b}_i,\mathbf{e_{i}}) + d(\tilde{b}_i,e_i) =\\
    & \quad \quad \quad \quad \quad \quad \quad \quad \quad \quad \quad \quad \quad \quad \quad  =\sum_{j\in \mathbb{S}_{i,\epsilon}\setminus i} (\tilde{b}_i-e_j)^2 + (\tilde{b}_i-e_i)^2 =\sum_{j\in \mathbb{S}_{i,\epsilon}} (\tilde{b}_i-e_j)^2 \\
    &D(\tilde{e}_i) = \underbrace{d(e'_i,b_i)}_{\text{authenticity}} + \beta_{\text{ampl}}\gamma(\tilde{e}_i, b_i) = (\tilde{e}_i-b_i)^2 +  \beta_{\text{ampl}}\gamma(\tilde{e}_i, b_i) \\
    &\gamma(\tilde{e}_i, b_i) = \begin{cases}
        (\tilde{e}_i-n)^2 \quad \quad \mbox{if}\;\;b_i>n/2\\
        0 \quad \quad  \quad \quad \; \quad \mbox{if}\;\;b_i=n/2\\
        (\tilde{e}_i)^2 \quad \quad \;\;\;  \quad \mbox{if}\;\;b_i<n/2
    \end{cases}\\
    &\beta_{pers}= \beta_{expr} = \infty
\end{align*}
It is evident how the amplification term pushes the belief expression towards the extreme the personal belief was leaning to. Indeed, the expressed belief update, i.e., the belief minimizing $D(\tilde{e}_i)$, is\footnote{the discrete analogous of (see \eqref{discrete minimum})}
\begin{equation}
    e_i^* = \begin{cases}
         \frac{b_i + \beta_{\text{ampl}}n}{1+\beta_{\text{ampl}}} \quad \;\; \mbox{if}\;\;b_i>n/2\\
         b_i \quad \quad  \quad \quad \; \;\;\mbox{if}\;\; b_i=n/2\\
         \frac{b_i}{1+\beta_{\text{ampl}}}\quad \;\;\; \;\; \mbox{if}\;\;b_i<n/2
    \end{cases}
\end{equation}
\\
\\
Considering for each agent $i=1,\ldots,N$ a potentially different private $x_i\in [0,1]$ and expressed $\hat{x}_i\in [0,1]$ beliefs, the agent's personal and expressed belief updates in the EPO model \cite{anderson2019recent} respectively read 
\begin{align}
    &x_i \;\;\leftarrow\;\; \lambda_i \bigg[a_{ii}x_i + \sum_{j\neq i} a_{ij}\hat{x}_j\bigg] + (1-\lambda_i)\bar{x}_i   \\
    &\hat{x}_i \;\;\leftarrow\;\; \phi_i x_i + (1-\phi_i)\hat{x}_{avg}
\end{align}
where $\hat{x}_{avg} = \frac{1}{N}\sum_{j=1}^N  \hat{x}_j$ and $\phi_i\in [0,1]\;\forall\;i$, $a_{ij}, \lambda_i, \bar{x}_i$ as in the DeGroot/FJ model. $\phi_i$ represents the individual's ability to withstand group pressure. $\hat{x}_{avg}$ is the total average public belief\footnote{Some versions of the model consider, instead of the total average belief, the local one perceived by the agent $i$, which is usually indicated with $\hat{x}_{avg,i}$.}. The FJ model is recovered for $\phi_i=1\;\forall \;i$; the DeGroot model is recovered for $\phi_i=\lambda_i=1\;\forall \;i$. Similarly to the FJ model, the discretized EPO is represented within the PES model by setting
\begin{align}
    &D(\tilde{b}_i) =  \underbrace{d(\tilde{b}_i,\mathbf{s_{(i)}})}_{\text{social influence}}+ \underbrace{d(\tilde{b}_i,b_i)}_{\text{inertia}} + \underbrace{d(\tilde{b}_i)}_{\text{context}}  \\
    &\underbrace{d(\tilde{b}_i,\mathbf{s_{(i)}})}_{\text{social influence}} = d(\tilde{b}_i,\mathbf{e_{i}}) =\lambda_i\sum_{j \neq i} a_{ij}(\tilde{b}_i-e_j)^2 \\   
    &\underbrace{d(\tilde{b}_i,b_i)}_{\text{inertia}} = \lambda_i a_{ii}(\tilde{b}_i-b_i)^2 \\ 
    &\underbrace{d(\tilde{b}_i)}_{\text{context}} = (1-\lambda_i)(\tilde{b}_i-\bar{b}_i)^2\\
    &\\
    &D(\tilde{e}_i) =  \underbrace{d(\tilde{e}_i,b_i)}_{\text{authenticity}} +  \underbrace{d(\tilde{e}_i,\mathbf{s_{(i)}})}_{\text{conformity}}  \\
    &\underbrace{d(\tilde{e}_i,b_i)}_{\text{authenticity}} = \phi_i(\tilde{e}_i - b_i)^2\\
    &\underbrace{d(\tilde{e}_i,\mathbf{s_{(i)}})}_{\text{conformity}} = d(\tilde{e}_i,\mathbf{e_{i}}) = (1-\phi_i)\frac{1}{N} \sum_{j=1}^{N} (\tilde{e}_i-e_j)^2\\
    &\\
    &\beta_{pers}= \beta_{expr} = \infty
\end{align}
From the perspective of the PES model, the SOAM and the EPO is a process of optimal dissonance reduction in the personal and expressed belief update. In the personal update, the dissonance comes from the dissimilarity with respect to the expressed belief of the neighbouring agents (social influence), and in the EPO also from a fixed preference (context effects) and from the previous personal belief of the focal agent (inertia). Regarding the expression, in the SOAM the dissonance results from the dissimilarity with the personal belief plus a penalty for expressing moderate beliefs (authenticity + amplification), in the EPO from the dissimilarity with the personal belief and with the interacting agent's expressed beliefs (authenticity + conformity). Both the models assumes perfect representations, given the belief expressions.

\begin{table}[H]
    \centering
    \scriptsize
    \begin{tabular}{|p{1.5cm}|p{1.5cm}|p{1.8cm}|p{5.7cm}|p{1.4cm}|}
    \hline
    \textbf{Model}  & \textbf{Spectrum} & \textbf{Processes} & \textbf{Dissonances} & \textbf{Beta}\\
    \hline
    \textit{All of these models} & $S_s$=$S_e$  & Perfect validity \newline No ego projection & $D(\tilde{s}_{j(i)}) = \underbrace{(\tilde{s}_{j(i)}-e_j)^2}_{\text{validity}}$ & $\beta_{soc}$=$\infty$ \\
    &&&&\\
    \hline    
    \textbf{\textit{Concealed Voter Model}} & $S_b$=$S_e$=$\{-1,1\}$ & Self-reinforcement \newline Authenticity \newline Conformity& $D(\tilde{b}_i) = \underbrace{(\tilde{b}_i - e_i)^2}_{\text{self-reinforcement}}$ \vspace{0.2cm}\newline   $D(\tilde{e}_i) = \begin{cases}
         \underbrace{(\tilde{e}_i - b_i)^2}_{\text{authenticity}} \\
        \underbrace{(\tilde{e}_i - e_j)^2}_{\text{conformity}}
    \end{cases}$  & $\beta_{pers}$=$\infty$ \vspace{0.9cm} \newline $\beta_{expr}$=$\infty$ \\
    &&&&\\
    \hline
    \textbf{\textit{2-layers Ising}} & $S_b$=$S_e$=$\{-1,1\}$ & Social influence \newline Self-reinforcement \newline Context \newline Authenticity \newline Conformity & $D(\tilde{b}_i) = \beta_{\text{socinf}}(\underbrace{- \tilde{b}_i \sum_{j\in \mathcal{N}_i} e_j}_{\text{social influence}}) +$ \newline $+ \beta_{\text{selfreinf}}\underbrace{(-  \tilde{b}_i e_i)}_{\text{self-reinforcement}} \underbrace{- h_i\tilde{b}_i }_{\text{context}} $ \vspace{0.3cm}\newline  $D(\tilde{e}_i) = \beta_{\text{conf}}(\underbrace{-\tilde{e}_i \sum_{j\in \mathcal{N}_i} e_j}_{\text{conformity}}) + \beta_{\text{auth}}(\underbrace{-\tilde{e}_i b_i}_{\text{authenticity}})$& $\beta_{pers}$ finite \vspace{1.8cm} \newline $\beta_{expr}$ finite\\
    &&&&\\
    \hline
    \textbf{\textit{Social Opinion Amplification Model (SOAM)}} & $S_b $=$S_e$=\newline =$\{0,1,2,\ldots,n\}$ & Social influence \newline Authenticity \newline Incentive for belief amplification & $D(\tilde{b}_i) = \underbrace{\sum_{j\in \mathbb{S}_{i,\epsilon}\setminus i} (\tilde{b}_i-e_j)^2 }_{\text{social influence}} + \underbrace{(\tilde{b}_i-e_i)^2}_{\text{self-reinforcement}}$ \newline \newline \newline $D(\tilde{e}_i) = \underbrace{(\tilde{e}_i-b_i)^2}_{\text{authenticity}} + \beta_{\text{ampl}}\underbrace{\gamma(\tilde{e}_i, b_i)}_{\text{amplification}}$ \vspace{0.03cm}\newline $\gamma(\tilde{e}_i, b_i) = \begin{cases}
        (\tilde{e}_i-n)^2 \quad \quad \mbox{if}\;\;b_i>n/2\\
        0 \quad \quad  \quad \quad \; \quad \mbox{if}\;\;b_i=n/2\\
        (\tilde{e}_i)^2 \quad \quad \;\;\;  \quad \mbox{if}\;\;b_i<n/2
    \end{cases}$ \vspace{0.03cm}  & $\beta_{pers}$=$\infty$ \vspace{1.2cm} \newline $\beta_{expr}$=$\infty$ \\
    &&&&\\
    \hline
    \textbf{\textit{Expressed-Private-Opinion (EPO)}} & $S_b $=$S_e$=\newline =$\{0,1,2,\ldots,n\}$ & Social influence \newline Inertia \newline Context \newline Authenticity \newline Conformity & $D(\tilde{b}_i) =  \lambda_i\underbrace{\sum_{j \neq i} a_{ij}(\tilde{b}_i-e_j)^2}_{\text{social influence}}+ \lambda_i\underbrace{ a_{ii}(\tilde{b}_i-b_i)^2}_{\text{inertia}} + $\newline $+(1-\lambda_i)\underbrace{(\tilde{b}_i-\bar{b}_i)^2}_{\text{context}}$ \newline \newline \newline $D(\tilde{e}_i) =  \underbrace{\phi_i(\tilde{e}_i - b_i)^2}_{\text{authenticity}} +  \underbrace{(1-\phi_i)\frac{1}{N} \sum_{j=1}^{N} (\tilde{e}_i-e_j)^2}_{\text{conformity}}$   & $\beta_{pers}$=$\infty$ \vspace{2cm} \newline $\beta_{expr}$=$\infty$ \\
    &&&&\\
    \hline
    \end{tabular}
    \caption{\textbf{Models with distinct expressed and personal beliefs} (\textit{table}) and \textbf{corresponding version of the PES model} (\textit{figure \ref{fig:PES_model exressed and personal}}).\\These models make no distinctions between social and expressed beliefs. Within the PES model, this is represented by considering a perfect representation of the beliefs of the interacting individuals, i.e., infinite contribution of the validity in the social dissonance.}
    \label{tab: models with distinct personal/ expressed}
\end{table}

\subsection{Models with distinct personal and social beliefs}
\begin{figure}[H]
    \centering
    \includegraphics[width=0.8\textwidth]{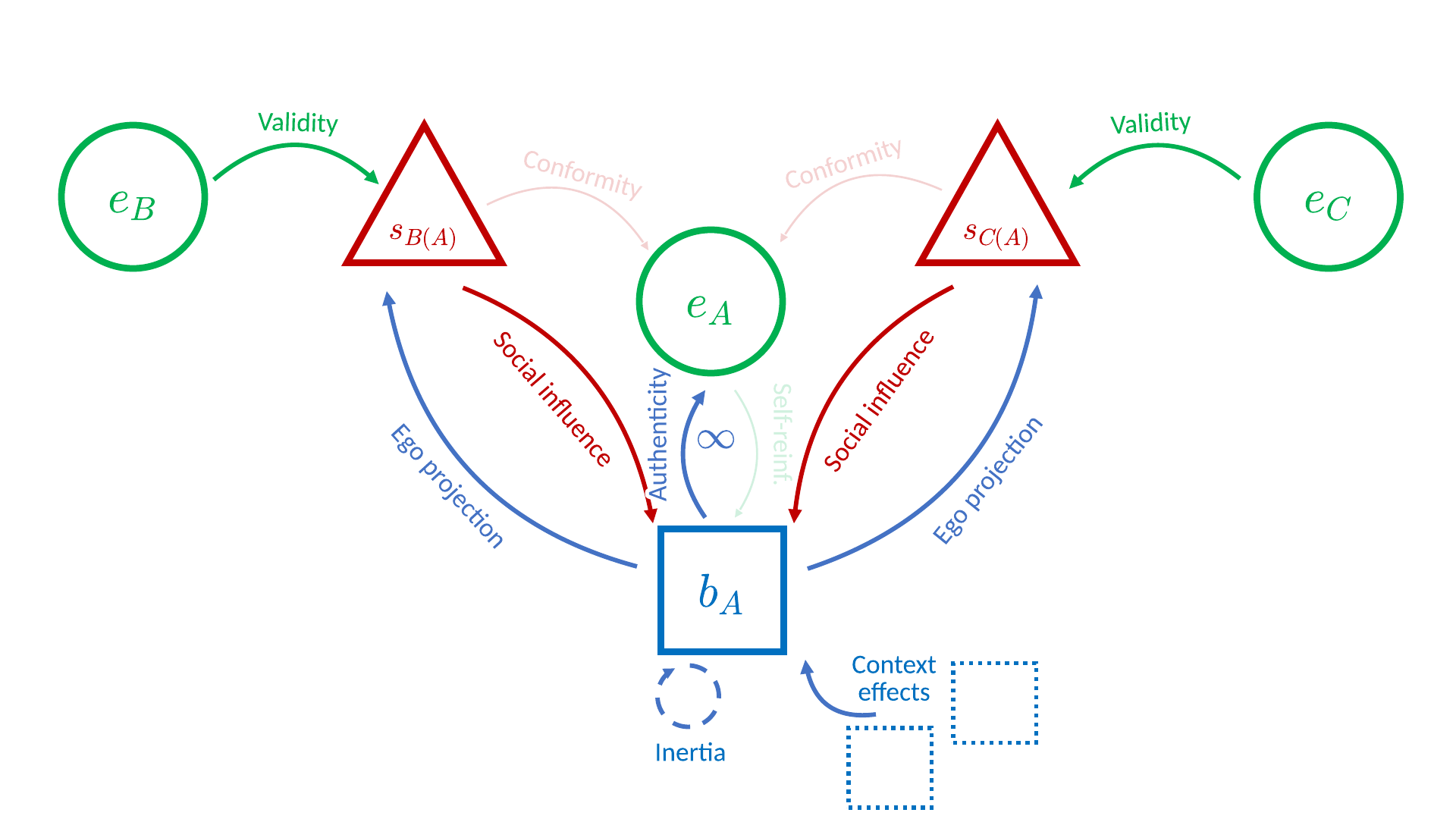}
        
    \vspace{0.5cm}
    \scriptsize
    \begin{tabular}{|p{1.3cm}|p{1.3cm}|p{1.65cm}|p{6.9cm}|p{1.2cm}|}
    \hline
    \textbf{Model}  & \textbf{Spectrum} & \textbf{Processes} & \textbf{Dissonances} & \textbf{Beta}\\
    \hline
    \textit{All of these models} & $S_e$=$S_b$  & Full authenticity \newline No conformity& $D(\tilde{e}_i) = \underbrace{(\tilde{e}_i-b_i)^2}_{\text{authenticity}}$ & $\beta_{expr}$=$\infty$\\
    &&&&\\
    \hline
    \textbf{\textit{NB theory with social beliefs}} & $S_b$ = $S_s $=$\newline  $=$\{-1,-.66,$\newline$-.33,0,.33,$\newline$.66,1\}$ & Social influence \newline Context \newline Validity \newline Ego projection& $D(\tilde{b}_i) = \beta_{soc,i}^{NB}\underbrace{\bigg[ -\tilde{b}_i\sum_{j\in \mathcal{N}_i} \rho_{ij} s_{j(i)}\bigg]}_{\text{social influence}} +$\newline$ + \beta_{pers,i}^{NB}\underbrace{\bigg[ -\tau_i \tilde{b}_i - \tilde{b}_i \sum_{a\in topics} w_i^{(a)} b_i^{*{(a)}}\bigg]}_{\text{context}}$  \newline \newline \newline $D(\tilde{s}_{j(i)}) = \beta_{ext,i}^{NB}\underbrace{\bigg[ -\alpha_{ij} \tilde{s}_{j(i)}b_j\bigg]}_{\text{validity}} + \underbrace{\beta_{soc,i}^{NB}\bigg[ - \rho_{ij} \tilde{s}_{j(i)}b_i\bigg]}_{\text{ego projection}}$ & $\beta_{pers}$=$1$  \vspace{2.4cm} \newline $\beta_{soc}$=$1$\\
    &&&&\\
    \hline
    \end{tabular}
    \caption{\textbf{Models with distinct social and personal beliefs} (\textit{table}) and \textbf{corresponding version of the PES model} (\textit{figure}).\\These models make no distinctions between personal and expressed beliefs. Within the PES model, this is represented by considering a fully authentic belief expression, i.e., infinite contribution of the authenticity in the corresponding dissonance.}
    \label{tab: models with distinct personal/social}
\end{figure}

There are not many belief dynamics models which consider possible distorsions in the representation of the interacting agent's beliefs. However, they do not distinguish between expressed and personal beliefs, i.e.,
\begin{equation}
     e_i = b_i  \quad \quad \quad \mbox{at any time }, \;\;\;\forall \;i
\end{equation}
The lack of possible distortions in the expression of the personal belief can be realized, within the PES model, by setting 
\begin{align}
    &D(\tilde{e}_i) = \underbrace{d(\tilde{e}_i,b_i)}_{\text{authenticity}} = (\tilde{e}_i-b_i)^2\\
    &\beta_{expr}=\infty    \label{eq distinct social and personal}
\end{align}
In this realization there is no distortion in belief expression, e.g. due to conformism. Figure \ref{tab: models with distinct personal/social} and the accompanying table show the PES model in this setting, along with the social-psychological processes of the \textit{Network of Beliefs (NB) theory, with social beliefs}.\\
\\
\textbf{NB theory with social beliefs}\\
The network of beliefs (NB) theory \cite{dalege2023networks} considers personal and social beliefs, both on a discrete set typically with values in the range $[-1,1]$. For example, in the simulations of the model, the authors of \cite{dalege2023networks} typically take
\begin{equation}
    S_b = S_s = \{-1,-0.66,-0.33,0,0.33,0.66,1\}
\end{equation}
As in the PES model, the principle for the update is the one of stochastic dissonance reduction. However, they consider a different implementation of the Glauber dynamics with respect to the PES model, including a dependency on the previous state which is absent in \eqref{eq Glauber dynamics} (unless in the form of the inertia): the probability of each possible update depends on the difference in the dissonance between the one with the potential update and the current one. The other important difference is that in NB theory one can write a total dissonance function for a focal individual $i$, which reads\footnote{Importantly, this does not mean that the model has a global potential to minimize (and can be reconducted to an Hamiltonian system), because the link between the focal agent's social belief and the interacting agent's personal belief is not reciprocated.}
\begin{equation}
\begin{split}
    D^{NB}_{tot,i}(b_i, \mathbf{s_{(i)}}) &= \beta_{pers,i}^{NB}\bigg[ -\tau_i b_i - b_i \sum_{a\in topics} w_i^{(a)} b_i^{*{(a)}}\bigg] +\\ 
    &+ \beta_{soc,i}^{NB}\bigg[ -b_i\sum_{j\in \mathcal{N}_i} \rho_{ij} s_{j(i)}\bigg] + \beta_{ext,i}^{NB}\bigg[ -\sum_{j\in \partial i} \alpha_{ij} s_{j(i)}b_j\bigg]
\end{split}
\end{equation}
where $\tau_i, w_{i}^{(a)}, \rho_{ij}, \alpha_{ij}$ $\forall i,j=1,\ldots,N$, and $a$ indexing each element in the set of non-focal beliefs, are exogenous parameters. Moreover, $\beta_{pers}^{NB^*}, \beta_{soc}^{NB^*}, \beta_{ext}^{NB^*}$ are interpreted as the attention, or weight, to respectively the personal/social/external dissonances (respectively the dissonance coming from internal, social and external validation processes) within the total dissonance function.

\noindent
Notice that multiplicative distances across all the beliefs are used: such choice, which systematically favours the update towards more extremist beliefs, is justified by showing that the simulations with the multiplicative distance better agree with data. Within the PES model, writing a general total dissonance of the agent is not possible because the links can be unreciprocated. However, apart from the rather marginal difference in the update, from the PES model we can reconduct to the NB theory with social beliefs through
\begin{align}
    &D(\tilde{b}_i) = \underbrace{d(\tilde{b}_i,\boldsymbol{s_{(i)}})}_{\text{social influence}} + \underbrace{d(\tilde{b}_i)}_{\text{context}}\\ \notag
    &\underbrace{d(\tilde{b}_i,\boldsymbol{s_{(i)}})}_{\text{social influence}} = \beta_{soc,i}^{NB}\bigg[ -\tilde{b}_i\sum_{j\in \partial i} \rho_{ij} s_{j(i)}\bigg] \\
    &\underbrace{d(\tilde{b}_i)}_{\text{context}} = \beta_{pers,i}^{NB}\bigg[ -\tau_i \tilde{b}_i - \tilde{b}_i \sum_{a\in topics} w_i^{(a)} b_i^{*{(a)}}\bigg]\\
    &\\
    &D(\tilde{s}_{j(i)}') = \underbrace{d(\tilde{s}_{j(i)},b_j)}_{\text{validity}} + \underbrace{d(\tilde{s}_{j(i)},b_i)}_{\text{ego projection}} \\ \notag
    &\underbrace{d(\tilde{s}_{j(i)},b_j)}_{\text{validity}} = \beta_{ext,i}^{NB}\bigg[ -\alpha_{ij} \tilde{s}_{j(i)}b_j\bigg]\\
    &\underbrace{d(\tilde{s}_{j(i)},b_i)}_{\text{ego projection}} = \beta_{soc,i}^{NB}\bigg[ - \rho_{ij} \tilde{s}_{j(i)}b_i\bigg]\\
    &\\
    &\beta_{pers} = \beta_{soc} = 1
\end{align}
Notice that, differently from $\beta_{pers}^{NB^*}, \beta_{soc}^{NB^*}, \beta_{ext}^{NB^*}$ in the NB theory, within the PES model, $\beta_{pers}, \beta_{soc}, \beta_{expr} $ can be interpreted as the attention to the maximization of the assigned utility function within the update of the corresponding belief, or as the \textit{rationality} of the agent in each of the updates.

\section{Comparison with the model of cultural evolution of Gavrilets et al. (2026) and other norm-utility models}\label{app:gavrilets}
The only class of models, to the best of our knowledge, that encompasses personal, expressed and second-order beliefs are the so-called \textit{norm-utility} models \cite{gavrilets2024modelling}, designed to reproduce the dynamics of social norms \cite{bicchieri2016norms, andrighetto2022research}. In these models, the utility function deciding the individual's action is composed of a material payoff and a psychological one, with personal and second-order beliefs entering the psychological side and evolving through generally DeGroot-like dynamics, or alternatively being treated as strategies and encompassed in an extended payoff matrix, as in \cite{zimmaro2026cultural}. In this sense, the action is rather a behaviour, with its own objective, external benefits and costs. However, the action can be interpreted also as a belief expression \cite{gavrilets2026cultural}. Under this second interpretation, and considering the model in Gavrilets et al. (2026) \cite{gavrilets2026cultural} because it is explicitly designed to reproduce and study pluralistic ignorance, we can map the evolving variables of \cite{gavrilets2026cultural} onto the ones of the PES model: specifically, their \textit{attitude} $y$ onto our personal belief $b$, their second-order belief $\tilde y$ about the average attitude in the population onto $s$, and their action $x\in\{-1,0,+1\}$, intended as the act of expressing one's attitude, onto our expressed belief $e$.

In \cite{gavrilets2026cultural}, the action maximizes $u_i(x)=Bx-c|x|+k_1y_ix+k_2\tilde y_ix$ (material payoff, cost of acting, cognitive dissonance, social influence; we write $B$ for their benefit $b$ and add a superscript $G$ to their parameters), while $y_i'=y_i+\alpha^G_1(x_i-y_i)+\beta^G_1|\bar x|(\bar x-y_i)+\gamma^G_1B$ and $\tilde y_i'=\tilde y_i+\alpha^G_2(y_i-\tilde y_i)+\beta^G_2(\bar x-\tilde y_i)+\gamma^G_2B$. Under best response ($\beta_{\text{pers,expr,soc}}\to\infty$), with $\mathcal S_b=\mathcal S_s=[-1,1]$, $\mathcal S_e=\{-1,0,+1\}$, and using that quadratic dissonances are minimized by the weighted average of the influencing beliefs and that the sum of squared distances reduces to the distance from the mean (see Sections below), the model is recovered by
\begin{align}
D(\tilde b_i) &= \underbrace{(1-\alpha^G_1-\beta^G_1|\bar e|)(\tilde b_i-b_i)^2}_{\text{inertia}} + \underbrace{\alpha^G_1(\tilde b_i-e_i)^2}_{\text{self-reinf.}} + \underbrace{\beta^G_1|\bar e|(\tilde b_i-\bar e)^2}_{\text{social infl.}^*} \underbrace{-2\gamma^G_1B\,\tilde b_i}_{\text{context}}, \nonumber\\
D(\tilde s_{j(i)}) &= \underbrace{(1-\alpha^G_2-\beta^G_2)(\tilde s_{j(i)}-s_{j(i)})^2}_{\text{inertia}} + \underbrace{\alpha^G_2(\tilde s_{j(i)}-b_i)^2}_{\text{ego proj.}} + \underbrace{\beta^G_2(\tilde s_{j(i)}-e_j)^2}_{\text{validity}} \underbrace{-2\gamma^G_2B\,\tilde s_{j(i)}}_{\text{context}^\dagger}, \nonumber\\
D(\tilde e_i) &= -u_i(\tilde e_i) = \underbrace{-k_1b_i\tilde e_i}_{\text{authenticity}} \underbrace{-k_2s_i\tilde e_i}_{\text{conformity}} \underbrace{-B\tilde e_i+c|\tilde e_i|}_{\text{ext. incentives/costs}^\dagger},
\end{align}
with $\tilde y_i\to s_i$ (the mean of the $s_{j(i)}$) and $\bar x\to\bar e$, provided $\alpha^G_1+\beta^G_1\le1$ and $\alpha^G_2+\beta^G_2\le1$. \\
\\
The two models thus share the same architecture, and most processes coincide: Gavrilets et al.'s \textit{social projection} and \textit{learning} are our ego projection and validity, and their \textit{cognitive dissonance} splits into two distinct PES processes, authenticity ($k_1$) and self-reinforcement ($\alpha^G_1$). Both models find that pluralistic ignorance is an out-of-equilibrium phenomenon and that the discreteness of the action/expression is key to its emergence. The main differences are \\
(1,$*$) social influence on attitudes comes directly from others' actions rather than being mediated by social beliefs, so that misperception affects what people do but not what they come to believe; its weight scales with $|\bar x|$, which has no clear cognitive meaning and cannot distinguish a silent population from a polarized one (both give $\bar x=0$);\\
(2,$\dagger$) PES has no external incentives on expressions, which are typical of norm-following behaviour rather than belief expression, and no context on social beliefs, a benefit-driven adjustment that in PES can emerge indirectly through ego projection;\\
(3) In general, PES generalizes \cite{gavrilets2026cultural} by allowing other discretizations of the expression spectrum, stochastic updates (which allow escaping metastable states, Fig.~\ref{fig:gamma1results}d), and different functional forms of the dissonances that allows a connection with a wide range of models of opinion dynamics.

\section{PES and pluralistic ignorance}\label{app:PI}
\begin{flushright}
\begin{minipage}{0.6\textwidth}
\raggedleft
\textit{``The universe is change;\\ our life is what our thoughts make it. ''}\\[4pt]
--- Marcus Aurelius, \textit{Meditations}, IV.3
\end{minipage}
\end{flushright}

\paragraph{Euclidean distance reduces to influence of the mean.}
Under a quadratic (Euclidean) dissonance the influence of the sum of dissonances from each member of a population coincides with the influence of the mean
under Glauber dynamics.
Indeed, writing
\begin{equation}
D(\tilde{\sigma}_i) = f(\tilde{\sigma}_i) + \frac{\beta}{N-1} \sum_{j=1,j\neq i}^N (\tilde{\sigma}_i - \sigma_j)^2
\label{dissonance general}
\end{equation}
defining $\bar{\sigma}_{-i}:= \frac{1}{N-1} \sum_{j=1,j\neq i}^N \sigma_j$ and $ \bar{\sigma}:=\frac{1}{N}\sum_{j=1}^N \sigma_j$, we have that
$$\sum_{j=1,j\neq i}^N (\tilde{\sigma}_i - \sigma_j)^2 = \sum_{j=1,j\neq i}^N (\tilde{\sigma}_i - \bar{\sigma}_{-i} +  \bar{\sigma}_{-i}-\sigma_j )^2=$$
$$= \sum_{j=1,j\neq i}^N \big[(\tilde{\sigma}_i - \bar{\sigma}_{-i})^2 + ( \bar{\sigma}_{-i}-\sigma_j )^2 - 2(\tilde{\sigma}_i - \bar{\sigma}_{-i})(  \bar{\sigma}_{-i}-\sigma_j )\big]=$$
$$= (N-1) (\tilde{\sigma}_i - \bar{\sigma}_{-i})^2 + (N-1)Var(\sigma_{-i}) $$
since the last term vanishes: $2(\tilde{\sigma}_i - \bar{\sigma}_{-i}) \sum_{j=1,j\neq i}^N (  \bar{\sigma}_{-i}-\sigma_j ) =0 $.\\
Thus, eventually
$$\frac{\beta}{N-1} \sum_{j=1,j\neq i}^N (\tilde{\sigma}_i - \sigma_j)^2 = \beta(\tilde{\sigma}_i - \bar{\sigma}_{-i})^2 + \beta Var(\sigma_{-i})$$
Now for the Glauber dynamics it holds that
$$\frac{e^{D(\tilde{\sigma}_i)}}{\sum_{\tilde{\sigma}_i\in S}e^{D(\tilde{\sigma}_i)}}  = \frac{e^{D(\tilde{\sigma}_i)-\beta Var(\sigma_{-i})}}{\sum_{\tilde{\sigma}_i \in S}e^{D(\tilde{\sigma}_i)-\beta Var(\sigma_{-i})}}$$
as $Var(\sigma_{-i})$ is independent on $\tilde{\sigma}_i$.\\
Thus, we can equivalently consider instead of \eqref{dissonance general}, 
$$D(\tilde{\sigma}_i) = f(\tilde{\sigma}_i) + \beta (\tilde{\sigma}_i - \bar{\sigma}_{-i})^2$$

\paragraph{Best response.}
With a dissonance function in the quadratic form (which is always the case with Euclidean distances, as proved in the previous paragraph)
\begin{equation}
    D(\tilde{\sigma}) = \sum_k \beta_k(\tilde{\sigma}-y_k)^2 
\end{equation}
the update of the focal belief under best response, i.e. the focal belief that minimizes the dissonance, reads
\begin{equation}
    \tilde{\sigma}^* = \frac{\sum_k \beta_k y_k}{\sum_k \beta_k}
\end{equation}

\paragraph{PES under reparametrization and best response}
The results above prove that the following dissonances 
\begin{align}
    \underbrace{D(\tilde{b}_i)}_{\text{personal}} &=\ 
       \underbrace{ (1-\gamma_i)  \eta_i \frac{1}{N-1}\sum_{j\neq i} (\tilde{b}_i-s_{j(i)})^2}_{\text{social influence}} + 
        \underbrace{ (1-\gamma_i)(1-\eta_i) (\tilde{b}_i-e_i)^2}_{\text{self-reinf.}} + 
        \underbrace{ \gamma_i (\tilde{b}_i- c^+)^2}_{\text{context}}  \nonumber\\[1ex]
    \underbrace{D(\tilde{e}_i)}_{\text{expressed}} &=\ 
       \underbrace{ \alpha_i(\tilde{e}_i-b_i)^2}_{\text{authenticity}} +  
        \underbrace{ (1-\alpha_i)\frac{1}{N-1}\sum_{j\neq i} (\tilde{e}_i-s_{j(i)})^2}_{\text{conformity}}  \label{eq: dissonances reparam} \\[1ex] 
    \underbrace{D(\tilde{s}_{j(i)})}_{\text{social}} &=\ 
        \underbrace{(1-\varphi_i)(\tilde{s}_{j(i)}-e_j)^2}_{\text{validity}} + 
         \underbrace{\varphi_i(\tilde{s}_{j(i)}-b_i)^2}_{\text{ego proj.}} \nonumber
\end{align}
produce the equations of the update of $s_i,e_i,b_i$ in the main text (Section \textit{Setup}), under best response and having considered the large $N$ approximation $\bar{\sigma}_{-i}\rightarrow \bar{\sigma}:=\frac{1}{N}\sum_{j=1}^N \sigma_j$.

\paragraph{No pluralistic ignorance at equilibrium, with continous spectra and homogeneous parameters.}
Let $\alpha,\varphi,\eta,\gamma\in[0,1]$ be homogeneous across individuals, and let the
spectrum be continuous. Assume $\alpha+\varphi>0$ and, if $\gamma=0$, also $\varphi<1$ and
$(\eta,\alpha)\neq(0,1)$. Then the only solutions of Eqs.~\eqref{eq:cs_s}--\eqref{eq:cs_b} are those with
$b_i=e_i=s_i=\bar b$ for all $i$, so that $PI=F=0$. If $\gamma>0$, then $\bar b=c^+$.

\noindent\textit{Proof.} At a fixed point,
\begin{align*}
s_i&=(1-\varphi)\,\bar e+\varphi\, b_i, \tag{\ref{eq:cs_s}}\\
e_i&=\alpha\, b_i+(1-\alpha)\, s_i, \tag{\ref{eq:cs_e}}\\
b_i&=(1-\gamma)\left[\eta\, s_i+(1-\eta)\, e_i\right]+\gamma\, c^+. \tag{\ref{eq:cs_b}}
\end{align*}

\emph{Averages.} Averaging \eqref{eq:cs_s} and \eqref{eq:cs_e} gives $\bar s=(1-\varphi)\bar e+\varphi\bar b$ and
$\bar e=\alpha\bar b+(1-\alpha)\bar s$. Substituting the second into the first,
\begin{equation*}
\left[1-(1-\varphi)(1-\alpha)\right]\bar s=\left[\varphi+(1-\varphi)\alpha\right]\bar b .
\end{equation*}
The two brackets are equal, and positive since $\alpha+\varphi>0$. Hence $\bar s=\bar b$,
and then $\bar e=\bar b$. Averaging \eqref{eq:cs_b} then gives $\gamma(c^+-\bar b)=0$, so $\bar b=c^+$
if $\gamma>0$.

\emph{Individual beliefs.} Using $\bar e=\bar b$, Eqs.~\eqref{eq:cs_s} and \eqref{eq:cs_e} give
\begin{equation*}
s_i-\bar b=\varphi\,(b_i-\bar b),\qquad
e_i-\bar b=\left[\alpha+(1-\alpha)\varphi\right](b_i-\bar b).
\end{equation*}
Subtracting $\bar b$ from both sides of \eqref{eq:cs_b} and using $\gamma(c^+-\bar b)=0$,
\begin{equation*}
b_i-\bar b=(1-\gamma)\left[\eta\,\varphi+(1-\eta)\left(\alpha+(1-\alpha)\varphi\right)\right](b_i-\bar b).
\end{equation*}
The bracket is a convex combination of $\varphi$ and $\alpha+(1-\alpha)\varphi$, both in
$[0,1]$, and it equals $1$ only if $\varphi=1$ or $(\eta,\alpha)=(0,1)$. Under the assumptions,
the coefficient on the right is therefore smaller than $1$, so $b_i=\bar b$. The two previous
identities then give $s_i=e_i=\bar b$. \hfill$\square$

\paragraph{No pluralistic ignorance at equilibrium, with continous spectra and heterogeneous parameters.}
Let $\alpha_i,\varphi_i,\eta_i\in(0,1)$ (for simplicity, we restrict to non-extreme parameters) and $\gamma_i\in[0,1]$,
possibly heterogeneous, and let the spectrum be continuous. Then the only solutions of
Eqs.~\eqref{eq:cs_s}--\eqref{eq:cs_b} are those with $b_i=e_i=s_i=\bar e$ for all $i$, so that $PI=F=0$.
If $\gamma_i>0$ for some $i$, then $\bar e=c^+$.

\noindent\textit{Proof.} At a fixed point,
\begin{align*}
s_i&=(1-\varphi_i)\,\bar e+\varphi_i\, b_i, \tag{\ref{eq:cs_s}}\\
e_i&=\alpha_i\, b_i+(1-\alpha_i)\, s_i, \tag{\ref{eq:cs_e}}\\
b_i&=(1-\gamma_i)\left[\eta_i\, s_i+(1-\eta_i)\, e_i\right]+\gamma_i\, c^+. \tag{\ref{eq:cs_b}}
\end{align*}
Subtracting $\bar e$ from both sides of \eqref{eq:cs_s} and \eqref{eq:cs_e} gives
\begin{equation*}
s_i-\bar e=\varphi_i\,(b_i-\bar e),\qquad
e_i-\bar e=\left[\alpha_i+(1-\alpha_i)\varphi_i\right](b_i-\bar e).
\end{equation*}
Subtracting $\bar e$ from both sides of \eqref{eq:cs_b} and substituting these two identities,
\begin{equation*}
\Big\{1-(1-\gamma_i)\big[\eta_i\varphi_i+(1-\eta_i)\big(\alpha_i+(1-\alpha_i)\varphi_i\big)\big]\Big\}\,(b_i-\bar e)
=\gamma_i\,(c^+-\bar e).
\end{equation*}
The square bracket is a convex combination of $\varphi_i$ and $\alpha_i+(1-\alpha_i)\varphi_i$,
both of which lie in $(0,1)$. Hence the curly bracket is positive, and $b_i-\bar e$, as well
as $e_i-\bar e$, equals a positive multiple of $\gamma_i(c^+-\bar e)$.

Averaging $e_i-\bar e$ over $i$ gives zero, so $(c^+-\bar e)$ times a sum of non-negative
terms, which is positive as soon as some $\gamma_i>0$, vanishes. Therefore either
$\gamma_i=0$ for all $i$, or $\bar e=c^+$. In both cases $\gamma_i(c^+-\bar e)=0$ for every $i$,
so $b_i=\bar e$, and the two identities above give $s_i=e_i=\bar e$. \hfill$\square$

\paragraph{Decomposition of pluralistic ignorance.}\label{si:PIcov}
Under best response and in the large-$N$ limit, each social belief reads (Eq.~\eqref{eq:cs_s} of the main text)
\begin{equation*}
s_i = (1-\varphi_i)\,\bar e + \varphi_i\, b_i .
\end{equation*}
Averaging over $i$, and using $\frac{1}{N}\sum_i \varphi_i b_i = \bar\varphi\,\bar b + \mathrm{Cov}(\varphi_i, b_i)$,
\begin{equation*}
\bar s = (1-\bar\varphi)\,\bar e + \bar\varphi\,\bar b + \mathrm{Cov}(\varphi_i, b_i).
\end{equation*}
Subtracting $\bar b$ and recalling $\mathrm{PI} = \bar s - \bar b$ and $F = \bar e - \bar b$ gives
\begin{equation}
\mathrm{PI} = (1-\bar\varphi)\,F + \mathrm{Cov}(\varphi_i, b_i).
\end{equation}
The identity holds at any time, since all social beliefs are refreshed after each update, and for any expression spectrum $S_e$, since only Eq.~\eqref{eq:cs_s} is used. For finite $N$, replacing $\bar e$ with $\bar e_{-i}$ adds a correction of order $1/N$. Under authentic expression ($F=0$) it reduces to Eq.~\eqref{eq:PI solo Cov} of the main text.

\paragraph{Simulations: Robustness}

\begin{figure*}[t]
    \centering
    \includegraphics[scale=0.17]{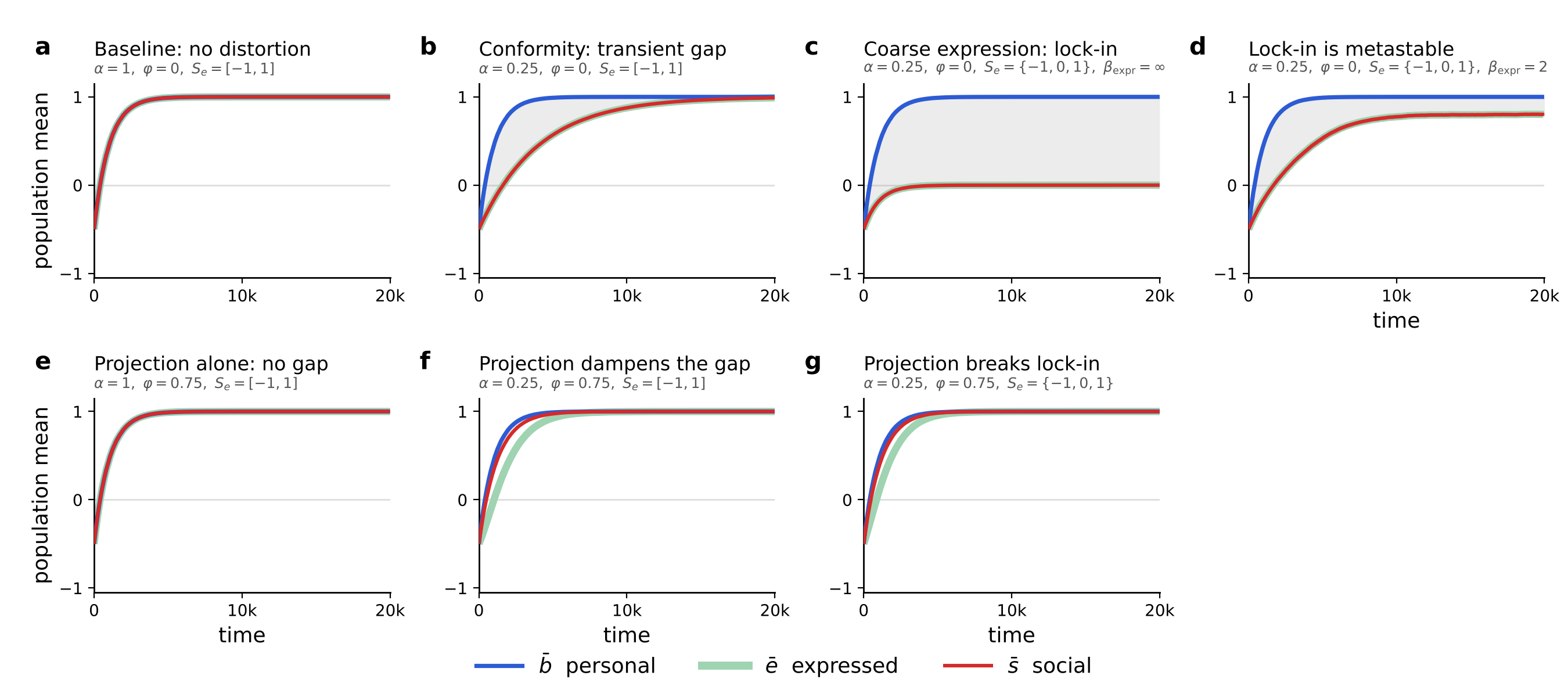}
    \caption{\textbf{Pluralistic Ignorance with $\gamma=1$.} Simulations for $\gamma=1, c^+=1$; $S_b=S_s=[-1,1]$, $\beta_{pers}=\beta_{soc}=\infty$, $\beta_{expr}=\infty$ unless specified, $\varphi, \alpha$ as in the legends of the subplots. Average personal, expressed and social beliefs of a population of $N=1000$ individuals, mean of $100$ runs (shading=$\pm$ 1 SD across runs). Initial beliefs normally distributed with $\langle b \rangle_0=-0.5$, $\sigma_b=0.05$. }
    \label{fig:gamma05results}
\end{figure*}

\begin{figure*}[t]
    \centering
    \includegraphics[scale=0.48]{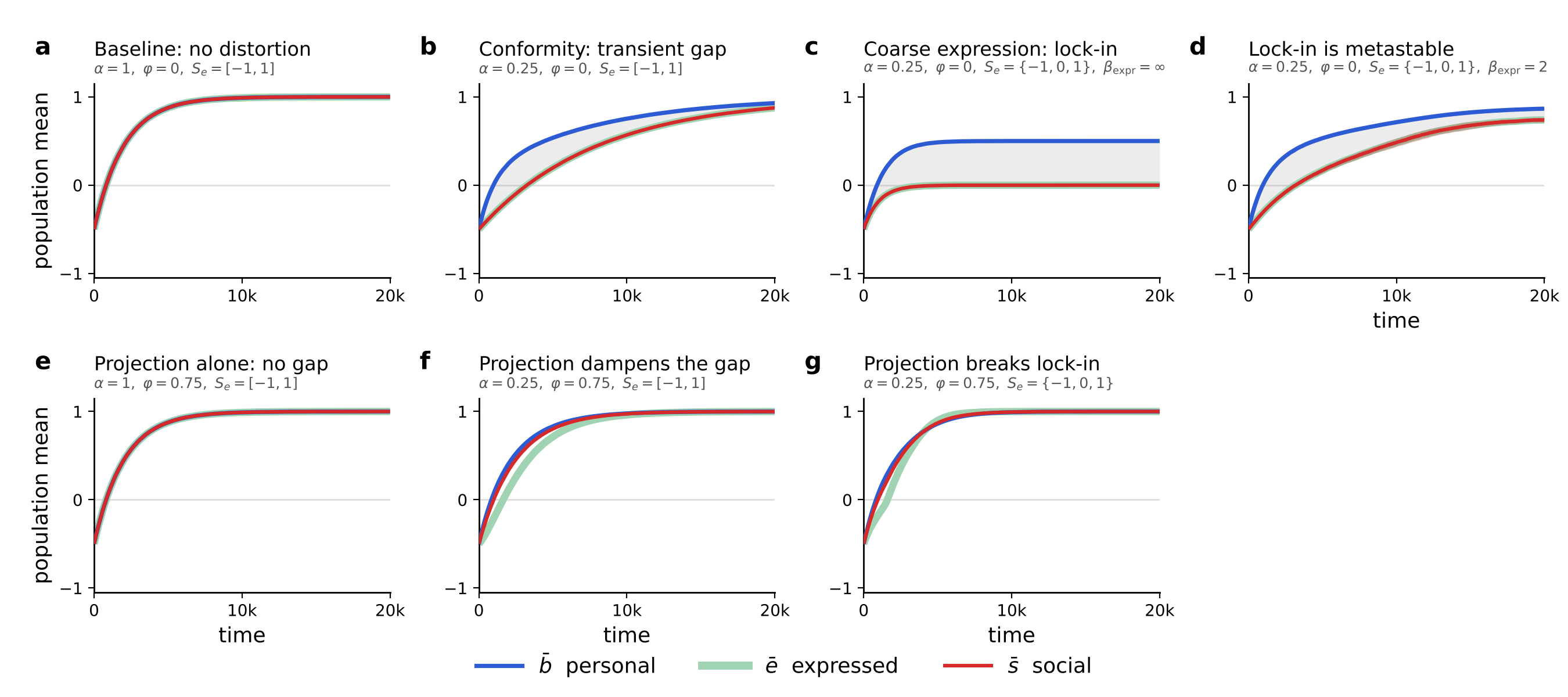}
    \caption{Same as Fig. \ref{fig:gamma05results}, except for $\gamma=0.5,\eta=1$. }
    \label{fig:gamma05eta1results}
\end{figure*}

\begin{figure*}[t]
    \centering
    \includegraphics[scale=0.48]{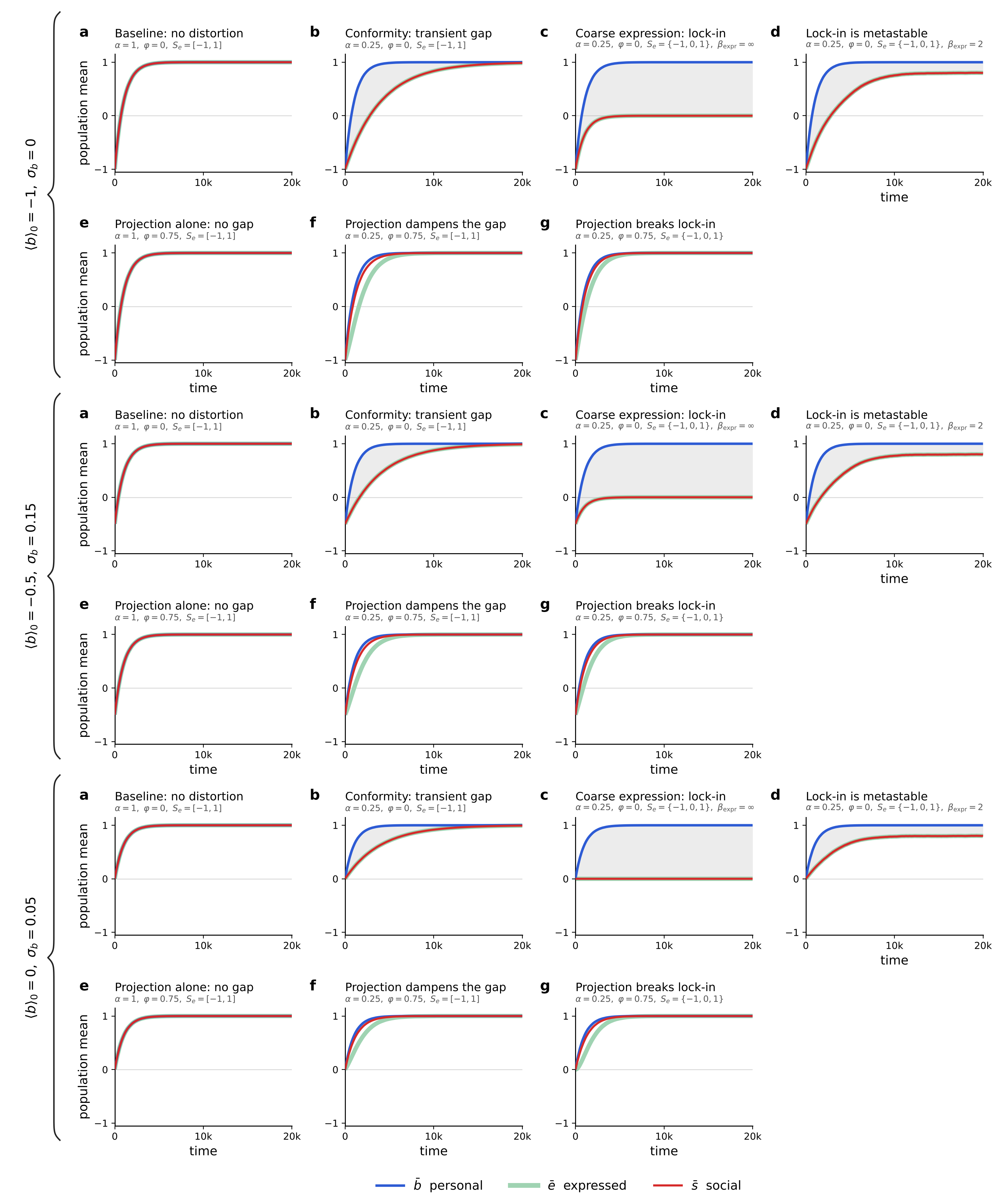}
    \caption{\text{Pluralistic ignorance: $\gamma=1$, varying initial beliefs (mean and dispersion as indicated on the left).}  }
    \label{fig:PIinitialbeliefs}
\end{figure*}

We test other combinations of $\gamma$ and $\eta$: $\gamma=1$ in Fig. \ref{fig:gamma05results}, $\gamma=0.5, \eta=1$ in Fig. \ref{fig:gamma05eta1results}. The pictures do not change qualitatively: the same mechanisms produce, respectively, a transient gap, a lock-in and its dissolution by ego projection. What changes is that the transitions are more abrupt especially for $\gamma=1$ and the overall level of pluralistic ignorance is higher. In Fig. \ref{fig:PIinitialbeliefs}, we change the initial conditions, varying the average personal belief of the population and its dispersion. The qualitative behaviour stays the same.

\end{document}